\documentclass[journal]{IEEEtran}

\usepackage{graphicx}
\usepackage[dvipsnames]{xcolor}
\usepackage{amssymb}
\usepackage{amsmath}
\usepackage{optidef}
\usepackage{array}
\usepackage{algpseudocode}
\usepackage{multirow}
\usepackage{algorithm}
\usepackage{cite}
\usepackage{csquotes}
\usepackage{float}
\ifCLASSINFOpdf
\else
\fi
\begin{document}
\title{Dependency Tasks Offloading and Communication Resource Allocation in Collaborative UAVs Networks: A Meta-Heuristic Approach}

\author{Loc X. Nguyen, Yan Kyaw Tun,~\IEEEmembership{Member,~IEEE,}
        Tri Nguyen Dang, Yu Min Park, Zhu Han,~\IEEEmembership{Fellow,~IEEE,}
        and Choong Seon Hong,~\IEEEmembership{Senior Member,~IEEE}% <-this % stops a space
\thanks{Loc X. Nguyen, Tri Nguyen Dang, Yu Min Park, Choong Seon Hong are with the Department of Computer Science and Engineering, Kyung Hee University, Yongin-si, Gyeonggi-do 17104, Rep. of Korea, e-mail: \{xuanloc088,trind,yumin0906,cshong\}@khu.ac.kr. }% <-this % stops a space
\thanks{Yan Kyaw Tun is with Teletraffic Systems, Division of Network and Systems Engineering, School of Electrical Engineering and Computer Science, KTH Royal Institute of Technology, Brinellvägen 8, 114 28 Stockholm, Sweden, and also with the Department of Computer Science and Engineering, Kyung Hee University,  Yongin-si, Gyeonggi-do 17104, Rep. of Korea, e-mail:{\{yktun\}@kth.se}.}
\thanks{Zhu Han is with the Electrical and Computer Engineering Department, University of Houston, Houston, TX 77004, and also with the Department of Computer Science and Engineering, Kyung Hee University, Yongin-si, Gyeonggi-do 17104, Rep. of Korea, email:{\{hanzhu22\}}@gmail.com}\vspace{-0.4in}
}

\maketitle

\begin{abstract}
In recent years, unmanned aerial vehicles (UAVs) assisted mobile edge computing systems have been exploited by researchers as a promising solution for providing computation services to mobile users outside of terrestrial infrastructure coverage. However, it remains challenging for the standalone MEC-enabled UAVs in order to meet the computation requirement of numerous mobile users due to the limited computation capacity of their onboard servers and battery lives. Therefore, we propose a collaborative scheme among UAVs so that UAVs can share the workload with idle UAVs. Moreover, current task offloading strategies frequently overlook task topology, which may result in poor performance or even system failure. To address the problem, we consider offloading tasks consisting of a set of sub-tasks, and each sub-task has dependencies on other sub-tasks, which is practical in the real world. Sub-tasks with dependencies need to wait for the resulting signal from preceding sub-tasks before being executed. This mechanism has serious effects on the offloading strategy. Then, we formulate an optimization problem to minimize the average latency experienced by users by jointly controlling the offloading decision for dependent tasks and allocating the communication resources of UAVs. The formulated problem appears to be NP-hard and cannot be solved in polynomial time. Therefore, we divide the problem into two sub-problems: the offloading decision problem and the communication resource allocation problem. Then a meta-heuristic method is proposed to find the sub-optimal solution of the task offloading problem, while the communication resource allocation problem is solved by using convex optimization. Finally, we perform substantial simulation experiments, and the result shows that the proposed offloading technique effectively minimizes the average latency of users, compared with other benchmark schemes.

\end{abstract}


% \fi
\IEEEpeerreviewmaketitle

\section{Introduction}
\IEEEPARstart{O}{ver} the last decade, the number of mobile users has grown exponentially, which eventually promotes the development in wireless communication and networking management to guarantee the quality of services (QoS) for all users \cite{mao2017survey}.
Along with that growth, computation-intensive applications such as virtual reality (VR), natural language processing, and fast navigation, which have become a necessary component of our lives.
Nonetheless, mobile users have difficulty catching up with the data processing due to the restrictions in terms of computing capacity and battery life.
Mobile cloud computing (MCC) with a large storage space, high processing speed, and unlimited energy resources relieves the pressure on mobile users by offering its resources \cite{marotta2015managing}, \cite{dinh2013survey}.
Nevertheless, as more and more mobile users appear, MCC is running into various problems, including high latency, poor coverage, weak security, and sluggish data transmission \cite{abbas2017mobile}.
Therefore, a new concept in the computing landscape has been invented to replace the outdated one: multi-access edge computing (MEC), which can resolve the above issues in the MCC system.
The term \enquote{edge computing} refers to data processing and storage that takes place at the \enquote{edge} of a network, close to the user \cite{yu2017survey}.
By doing this, we can drastically lower the transmission latency due to the fact that MEC servers are substantially closer to the users than cloud servers and also avoid the peak in traffic flows. Normally, we deploy the computing servers at the edge of the network, such as cellular base stations (BSs) or wireless access points (APs).

In the case of user devices operating in rural areas, such as mountains, forests, deserts, or underwater environments in temporary events, e.g. rescue operations in disaster locations and military operations, installing a new BS or APs requires lots of time and is a waste of money for one time used. Therefore, \cite{tun2020energy,cheng2019space,zhou2018uav,zhang2019joint,zhang2018stochastic} have proposed ideas that were using unmanned aerial vehicles (UAVs), such as drones, high-altitude platforms, and balloons, as communication platforms for providing service to mobile users. These vehicles are flexible to deploy in any special area and can also be reused. Moreover, MEC-enabled UAVs can collect data and perform computing tasks for those devices that do not have any direct access to terrestrial BSs or APs \cite{cheng2018air}.
In addition, MEC-enabled UAV servers can be used as relay stations for efficiently expanding the communication coverage.

Even though the MEC-enabled UAV system has great potentials and is expected to provide computing services better and faster than conventional-fixed location-MEC systems, there are some challenges that still stand in the way before adopting this technique in the real world.
First of all, the MEC-enabled UAV system has lower computing capacities and energy resources when compared to the cloud server or even traditional MEC system, and therefore it is eager to find an optimal offloading strategy for MEC-enabled UAV systems to lower the task latency under such limited conditions \cite{you2016energy,tun2019energy}.
Secondly, each MEC-enabled UAV system has to support numerous devices over such a broad coverage area while having fewer bandwidth resources to allocate compared to the communication resources of the BS-enabled MEC system.
Therefore, an optimization algorithm is required to allocate bandwidth resources among users instead of distributing them equally.
Moreover, the server attached to the UAV has much lower computing resources compared to the cloud server and the servers attached to the terrestrial BSs.
As a consequence, the task from users may fail to finish on time even only part of it was offloaded. Therefore, collaboration among UAVs has been proposed as a key to overcoming the task’s deadline and also boosting its energy efficiency.

Task offloading has been one of the main features in edge computing scenarios, which decides how the task will be executed.
In most cases, the task from a user is parameterized by two numbers: the input size and the number of CPU cycles required to finish the task.
By doing this, they can easily decompose the task into small sub-tasks, offload them to edge devices, and enable parallel execution on both edge and mobile devices. 
However, the dependency of the sub-tasks might have a significant influence on this parallelism \cite{han2019efficient}.
The dependency among sub-tasks can be simply understood as the start of a sub-task only happens when all the predecessor sub-tasks of it finish, because it requires the output information of the predecessor sub-tasks \cite{fan2019cost}.

% A sub-task may have predecessor sub-tasks that must be completed before the beginning of that sub-task, and it may also have successor sub-tasks, which cannot be executed until the task is completed [15].

Therefore, these dependency sub-tasks have to be executed sequentially instead of in parallel.
A directed acyclic graph (DAG) can often be used to demonstrate the dependencies among sub-tasks.
Each circle denotes the sub-task of the whole task, and the directed edge represents the topology among them, and each sub-task has a different computing requirement, result requirement, and input length of its.
As a result, the scheduling decision will become much more complicated, and it appears to be an NP-hard problem.

Recently, most user applications consist of a series of sub-tasks. For example, in \cite{jain2011handbook} the authors discussed about the process of face recognition task, which comprises four sub-tasks: face detection, face alignment, feature extraction, and finally feature matching. These sub-tasks have to be performed sequentially step by step due to the dependency among them. The output from the face detection: face location, size, and pose will be used as the input for the face alignment sub-task. The aligned face goes into the feature extraction module to get a feature vector. Then that feature vector will be compared with the features in the database to find the best match and output the face ID in the final sub-task. This is just a simple example of the dependency inside of the task, and there are still a lot of tasks that we have to consider the dependency inside.

Therefore, in this paper, we investigate the effects of offloading the tasks with dependencies on collaborative UAVs networks. The main contributions of this article are summarized as follows.
\begin{itemize}
 \item First, we propose the task model, which includes multiple sub-tasks and the dependency among them to capture the task topology in the real scenario. Then these tasks from users will be offloaded to MEC-enabled UAVs networks. The BS is the central controller and will decide the offloading strategy based on its knowledge of the tasks and the networks.

\item Secondly, we mathematically formulate the optimization problem of the proposed system model with the objective of minimizing the average latency of the mobile users by controlling the offloading decision and the communication resources allocated to each user. The formulated problem has been proven to be NP-hard.
\item Thirdly, we decompose the problem into two sub-problems: i) dependent task offloading decision problem, and ii) bandwidth allocation for users problem. Then we propose D-WOA, which is a meta-heuristic approach to solve the decision problem, which cannot be solved in polynomial time by traditional methods. The problem of allocating communication resources is convex, and so we can use the splitting conic solver (SCS) in CVXPY to solve it.
\item Finally, we conduct a series of simulations to evaluate the efficiency of the proposed method and compare it against benchmark schemes such as the exhaustive search algorithm and associated UAVs. We also show how the restricted energy of the UAVs affects the execution latency of the tasks.
\end{itemize}

The remainder of this paper will be organized as follows. In Section \ref{Related}, we will briefly discuss about related works. Then, we describe the system model and problem formulation in more detail in Section \ref{System}. Our proposed solution will be demonstrated in-depth in Section \ref{Proposed}. We show the simulation results and, based on those results, and draw a conclusion in Sections \ref{Performance} and \ref{Conclusion}, respectively.

\section{Related works}\label{Related}

\subsection{Multi-Access Edge Computing}
This subsection will discuss a summary of the literature on multi-access edge computing \cite{yang2018mobile,sun2019joint,li2020joint,xia2021data,liu2020data} . The authors in \cite{yang2018mobile} proposed artificial fish swarm algorithm (AFSA) to solve the task offloading problem with the assistance of femto relay BSs, their objective is to find the access and offloading decisions that return the least energy consumption. The authors in \cite{sun2019joint} wanted to maximize the sum of computing efficiency by a traditional optimization approach: the iterative gradient descent method. A two-stage heuristic optimization algorithm was proposed to allocate the communication resources and offload the computing tasks from multiple users to multiple servers using the least amount of energy in \cite{li2020joint}. The authors in \cite{xia2021data} proposed DUPA$^{3}$ game to allocate data, users, and power for caching problems in MEC to maximize the total data rate and the number of served users. The maximization of the total data rate leads to a low latency network, which is beneficial for mobile users, in \cite{liu2020data} the authors considered the data caching problem from the services provider's view. They proposed an approximation approach to solving the data caching problem with the goal of optimizing service providers' revenue.

\subsection{MEC-enabled UAVs Networks}
The MEC-enabled UAV has been received lots of attention from research scientists for its many advantages \cite{tun2020energy,yu2020joint,zhang2020energy,tun2022collaboration,seid2021collaborative}. In \cite{tun2020energy}, Yan \textit{et al.} proposed block successive upper-bound minimization (BSUM) approach to minimizing the energy consumed by IoT devices and UAV-aided MEC systems by finding the optimal offloading decision, resource allocation, and UAV route. The authors in \cite{yu2020joint} formulated a problem to minimize the combination value of service and UAV energy consumption by controlling the UAV coordinate, computation resource, communication resource, and finally, the splitting ratio of the task. Then the successive convex approximation based algorithm was proposed to find a sub-optimal solution. In \cite{zhang2020energy}, Zhang \textit{et al} optimized the offloading decision to minimize the cost of time and energy consumed by the system while considering the limited energy of the UAV-aided assisted MEC, then they proposed a game theory-based solution to find the offloading decision. In the works mentioned above, they considered UAVs operating independently from each other in the MEC network. The researchers in \cite{tun2022collaboration} and \cite{seid2021collaborative} proposed the collaboration among MEC-enabled UAVs in task computing and proved the effectiveness of it. In \cite{tun2022collaboration}, the authors optimized the offloading decision, the computing resource, and the communication resource under the energy limitation of computing participants to obtain the lowest latency of all the tasks. Compared with the previous paper, the collaboration among MEC-enabled UAVs offers more destinations for the task to be offloaded and therefore takes the pressure off the associated UAV. The authors in \cite{seid2021collaborative}, Seid \textit{et al.} proposed CCORA-DRL for each MEC-enabled UAV to learn an efficient computation offloading strategy and therefore acquire minimum service latency.
\subsection{Offloading Dependent Tasks}
This subsection will provide a summary of the literature on offloading dependency tasks \cite{fan2019cost,sundar2018offloading,shu2019multi,liu2020dependency,zhao2021offloading,sahni2020multihop}. Sundar \textit{et al.} in \cite{sundar2018offloading} was the first to study scheduling and offloading decision for tasks comprising dependency sub-tasks in a MCC system. They aimed to minimize the task execution cost subject to the task finish deadline by using time allocation with greedy scheduling (ITAGS) to effectively solve the proposed NP-hard problem. While the work in \cite{shu2019multi} considered the edge servers to execute the dependent tasks. The formulated problem became more complicated compared to the MCC system due to the heterogeneous computing resources of the edge servers and the requirement for information exchange among sub-tasks. They came up with the distributed earliest-finish time offloading algorithm to effectively reduce the latency of the dependency task from IoT devices. In \cite{liu2020dependency}, the dependency task was generated by the vehicular edge computing, and they considered minimizing the execution latency of the tasks by optimizing the offloading strategy for each sub-task: local execution (VEC) or offloading to the roadside unit. Then they proposed multiple applications multiple tasks scheduling (MAMTS) algorithm to calculate the sub-task prioritization and schedule the sub-task according to the prioritization. The researchers in \cite{fan2019cost} introduced a new metric to evaluate the efficiency of offloading strategy for dependency tasks and then proposed a heuristic offloading solution to shrink the cost of the process. In \cite{zhao2021offloading}, the author formulated offloading dependent tasks with service catching problem with the goal to minimize the makespan of the tasks and then designed a convex programming-based approach to tackle the problem. The authors in \cite{sahni2020multihop} jointly optimize the offloading decisions and the network flow in the collaborative edge computing systems to minimize the completion time of the tasks. Most of these works assumed that the edge server had a large computing capacity and infinite power resources. However, this assumption will not hold if we employ MEC attached to the UAV. These works either did not consider the reduction in computation resource for one sub-task when there were multiple sub-tasks being offloaded at the same time or assumed the edge executed one sub-task at one time like in \cite{sahni2020multihop}.

% Note that the IEEE does not put floats in the very first column
% - or typically anywhere on the first page for that matter. Also,
% in-text middle ("here") positioning is typically not used, but it
% is allowed and encouraged for Computer Society conferences (but
% not Computer Society journals). Most IEEE journals/conferences use
% top floats exclusively. 
% Note that, LaTeX2e, unlike IEEE journals/conferences, places
% footnotes above bottom floats. This can be corrected via the
% \fnbelowfloat command of the stfloats package.
\section{System model and Problem Formulation}\label{System}
\subsection{Overview System Model}

For simplicity, we leave out the transmit latency of sub-tasks in the figure, but in the problem formulation, we do include it.

\begin{figure}
    \centering
    \includegraphics[width=.5\textwidth]{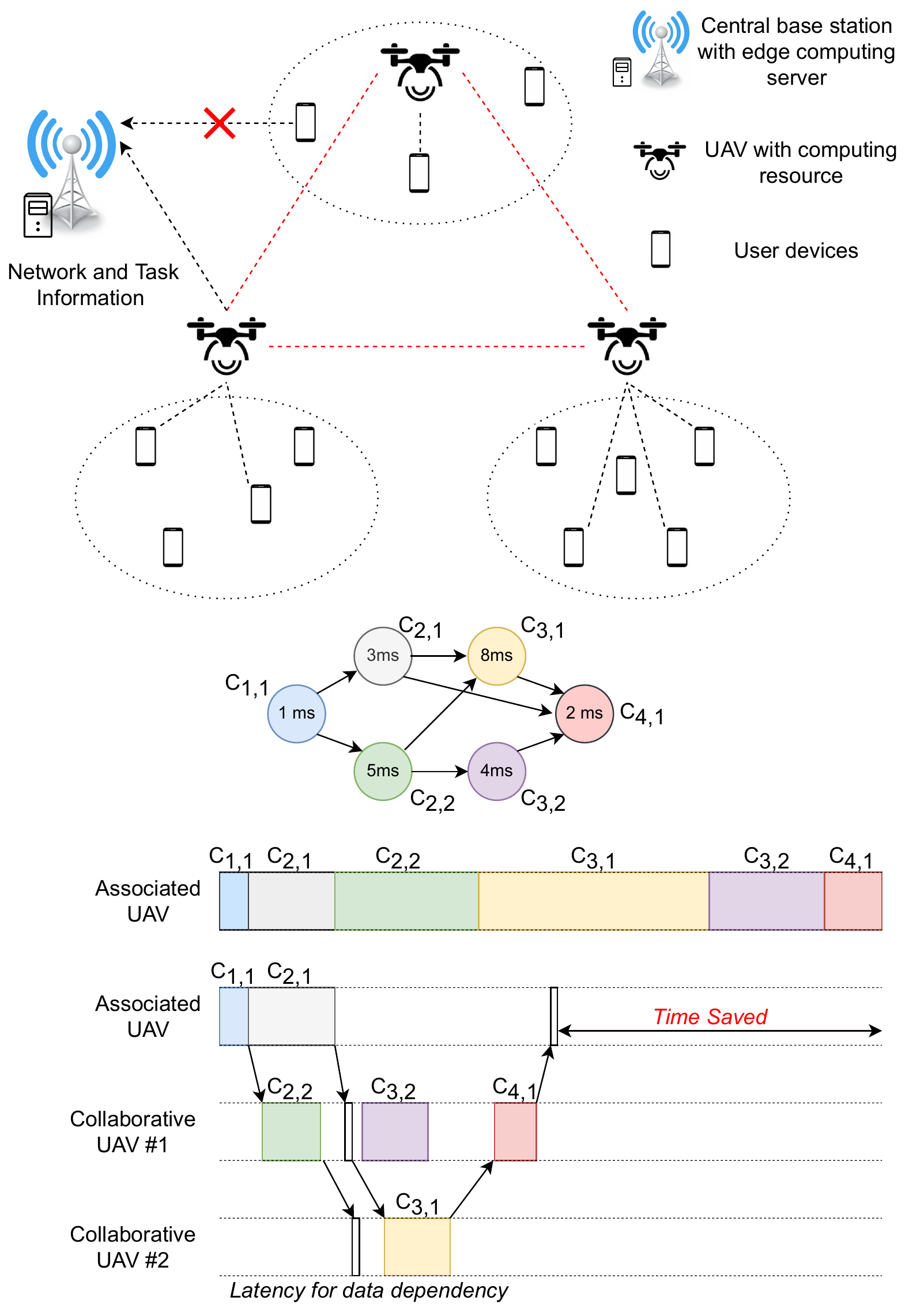}\\
    \caption{System model.}
    \label{fig:my_label}
\end{figure}

\textit{Network Model:} As shown in Fig. 1, we have a set $\mathcal{V}$ of $V$ MEC-enabled UAVs to provide communication and computation services to users in a certain area, and they can connect with the terrestrial BS with a computing server by both line of sight and non-line of sight networks.
In this paper, we make the assumption that there is no direct communication link between mobile users and the BS, or that users are not within the BS's service region.
The BSs' function is to serve as a centralized controller with collective knowledge about the network, which is in charge of selecting UAVs to offload sub-tasks and resource allocation.
Even if the direct connection does not exist, the BS can still gather information about each user's task through the third party-UAVs. To reduce the complexity of the problem, we assume that users are already assigned to serving UAVs determined by the channel quality. In particular, each UAV
$v \in \mathcal{V} $ knew its connected set of users, $\mathcal{U}_{v}$. The total number of mobile users in our network is defined as follows 
$\bigcup\limits_{v=1}^{V} \mathcal{U}_{v}$, where there is no user repetition among difference set. 

\textit{Task Model}: At a time, each user only can generate a single task. As a result, we occasionally refer to a task as being created by user $u$ by calling it task $u$. User $u \in \mathcal{U}_{v}$ has different requirements and structure $D_{u}$ that can be modeled as directed acyclic graph (DAG) $D_{u}=(T_{u},P_{u})$, where $T_{u}$ is the set of dependent sub-tasks in task $u$,
$T_{u}=\{j|1\leq j \leq N_{u}\}$, $P_{u}$ is a set of dependencies between
sub-tasks in task $u$ and $N_{u}$ is the total number of sub-task in task $u$. The computation requirement to finish sub-task $j$ of task $u$ depends on its input size, which is denoted as $H_{u,j}$. The amount of data dependency between sub-tasks $p$ and $j$ is denoted as $D^{u}_{p,j}$. If the preceding sub-tasks of sub-task $j$ are located in different devices, we have to transmit $D^{u}_{p,j}$ to the device which is responsible for sub-task $j$ before we start sub-task $j$. From now on, the set of predecessors and successor sub-tasks of sub-task $j$ are defined as $Pd_{u,j}$ and $Sc_{u,j}$, respectively. Besides, there are some sub-tasks that we can do simultaneously because they are independent of each other. For example, sub-tasks $c_{2,1}$ and $c_{2,2}$ can be processed at the same time, they can be thought of as co-sub-tasks. The time that user $u$ creates the task is called the release time, $Trel_{u}$.
\begin{table*}[htbp]
	%	\caption{Summary of Key Notations}	
	\caption{Summary of Notations}
	\renewcommand\arraystretch{1}
	\begin{center}
		\begin{tabular}{|m{1.8cm}|m{6cm}||m{1.8cm}|m{6cm}|}
			%		\begin{tabular}{|p{1.5cm}|p{6.5cm}||p{2cm}|p{6.5cm}|}
			\hline
			\hfil \textbf{Notation} & \hfil \textbf{Definition} & \hfil \textbf{Notation} & \hfil \textbf{Definition} \\ \hline \hline
			\hfil $\mathcal{V}$ & Set of UAVs, $|\mathcal{V}|$= V & \hfil $\mathcal{U}$ & Set of users, $|\mathcal{U}|$= U \\ \hline
			\ \hfil $\mathcal{D}_u$= ($\mathcal{T}_u$,$\mathcal{P}_u$)& Directed acyclic graph of task $u$, where $\mathcal{T}_u$ is the set of dependent sub-tasks and $\mathcal{P}_u$ is the set of edges & \hfil  $\mathcal{D}^{u}_{p,j}$ & The size of dependent data between sub-task \textit{p} and \textit{j} of task $u$ \\ \hline			
			
			\ \hfil  $N_{u}$ &  The total number of sub-tasks in task \textit {u} & \hfil $I_{v}$ & The information set of UAV $v$ \\ \hline					
			
			\ \hfil  $S_{u,j}$ &  The set of successor sub-tasks of sub-task \textit{j} in task \textit {u} & \hfil $P_{u,j}$ & The set of predecessor sub-tasks of sub-task \textit{j} in task $u$ \\ \hline			
			\ \hfil  $ H_{u,j}$ & The input size of sub-task \textit{j} of task $u$ & \hfil $Trel_{u}$ & The release time of the task from user $u$ \\ \hline
			\ \hfil $t^\textrm{load}_{u}$ & The transmission latency of task $u$ from the user to its associate UAV & \hfil $t^\textrm{dep}_{u,p,j}$ &  Latency for transmit dependent data from device execute sub-task \textit{p} to device execute sub-task \textit{j}  \\ \hline
			\ \hfil  $FT_{u,j}$     & The finish time of sub-task \textit{j} of task $u$   & \hfil  $ST_{u,j}$  & The start time of sub-task \textit{j} of task $u$ \\ \hline
			\ \hfil   $t^\textrm{exe}_{u,j}$    &  Execution latency of sub-task \textit{j} of task $u$  & \hfil  $C_{u}$  &   The number of CPU cycles required to execute one bit data   \\ \hline
			\ \hfil   $\Gamma^{u \rightarrow v}$    &  The spectrum efficiency from user $u$ to its associate UAV & \hfil  $P_{u}$  & The transmit power of user $u$  \\ \hline
			\ \hfil    $G_{v,u}$   &  The achievable channel gain between the user and UAV  & \hfil  $\sigma^{2}$  &  Gaussian noise power $n$    \\ \hline
			\ \hfil    $d^{u,v}$   &  The distance from user to its associated UAV & \hfil  $R^{u\rightarrow v}$  &  The achievable uplink transmission rate  \\ \hline
			\ \hfil    $\beta^{v}_{u}$   & The percentage of bandwidth will be allocated to user $u$  & \hfil  $B^{v}$  & Total accessible bandwidth that UAV $v$ can allocate to its users  \\ \hline
			\ \hfil    $t^{u\rightarrow v}$   & Data transmission latency from user to UAV & \hfil  $E^{v,\textrm{up}}$  &  Energy consumed by user to transmit task $u$ to UAV $v$ \\ \hline
			\ \hfil    $\theta^{u\rightarrow v}_{u,j}$   &  Decision variable whether sub-task $j$ of user $u$ is offloaded to its associate UAV $v$ or not & \hfil  $\gamma^{v\rightarrow w}_{u,j}$  & Binary variable which decide sub-task $j$ will be offloaded to the collaborated UAV \textit{w} in the network   \\ \hline
			\ \hfil    $f^{v}_{u}$   & The computing capacity of the MEC server of associated UAV   & \hfil  $F^\textrm{max}_{v}$  & The maximum computing resource of UAV $v$  \\ \hline
			\ \hfil    $t^{v,\textrm{exe}}_{u,j}$ & The amount of latency to execute the sub-task at UAV $v$   & \hfil  $E^{v,\textrm{exe}}_{u,j}$  & Energy consumption for execute sub-task \textit{j} \\ \hline
			\ \hfil    $R^{v\rightarrow w}$   &  Achievable data rate from UAVs $v$ to \textit{w}  & \hfil  $P_{v}$  &  The power that UAV $v$ use for transmission \\ \hline
			\ \hfil    $G^{v,w}$   & The channel gain between 2 collaborative UAVs   & \hfil  $d^{v,w}$  & The distance between UAVs $v$ and \textit{w}       \\ \hline
			\ \hfil    $E^{v\rightarrow w}_{u,j}$   & Energy consumption required to forward sub-task \textit{j} to UAV \textit{w}  & \hfil  $E^{\textrm{tol}}_{v}$  &  The total energy consumed of UAV $v$     \\ \hline
			\ \hfil   $P^{v,\textrm{hov}}$    &  The hovering power of UAV $v$  & \hfil   $t^\textrm{exe}_{u,j}$    &  Execution latency of sub-task \textit{j} of task $u$  \\ \hline
			\ \hfil    $E^{v,\textrm{hov}}$   & Hovering energy of UAV $v$  & \hfil  $t^{v,\textrm{hov}}$  & Hovering time of UAV $v$      \\ \hline
			
			\ \hfil   $t^{v\rightarrow b}$    &  The transmit latency from UAV $v$ to the BS  & \hfil  $R^{v\rightarrow b}$  &   The communication data rate from the user to the BS\\ \hline
			\ \hfil   $B^{v\rightarrow b}_{\textrm{mm}}$    &  The amount of bandwidth in mmWave frequency band allocated to UAV $v$  & \hfil  $P_{b,v}$  &   The energy received at the BS\\ \hline
			
			\ \hfil   $P^{v\rightarrow b}$    &  The power that UAV $v$ use to transmit information to the BS  & \hfil  $G^{tx}_{v}$  &   The antenna gain of UAV $v$\\ \hline
			\ \hfil   $G^{rx}_{b}$    &  The antenna gain of the BS  & \hfil  $d^{b}_{v}$  &   The distance from the BS to UAV $v$\\ \hline
			
			\ \hfil   $B^{\textrm{mm}}_{c}$    &  The carrier freqency of the mmWave link  & \hfil  $E^{v\rightarrow b}$  &   The energy consumption for transmit information from UAV $v$ to the BS\\ \hline

% 			\ \hfil    $$   &    & \hfil  $$  &       \\ \hline
		\end{tabular}
		\label{tab1}
	\end{center}
\end{table*}

In this paper, we assume the task is created by the local user, and to capture this assumption, we need to insert one dummy sub-task into the task, which has zero execution time and zero communication cost. Because of this, it is added at the start of the task without any changes in the original task DAG. The number of sub-task is increased to $N_{u}'$, where $N_{u}' = N_{u} + 1.$
The devices of mobile users have low computing capacity and energy, and so these limitations yield challenges for them to process the task quickly. Therefore, users offload the task to the associated UAVs in collaborative UAVs networks, which have a larger computing capacity via wireless links to finish the task within the minimum amount of time. To explain how we can do that, we will discuss the task and network models in the next subsection.
\subsection{Constraint of Task Model and Network Model}
\subsubsection{Task model}
As mentioned before, a dummy sub-task is added to ensure the task is created at user $u$. The start time of the dummy sub-task is set as the release time of the whole task, and since it does not require any computation resource, its finish time of it is defined in \cite{sahni2020multihop} as below:
\begin{equation}
    X_{uN_{u}'}=1 ,\ \forall u \in \mathcal{U}_{v},
\end{equation}
\begin{equation}
   ST_{uN_{u}'}= Trel_{u}, \ \forall u \in \mathcal{U}_{v},
\end{equation}
\begin{equation}
   FT_{uN_{u}'}= Trel_{u},\ \forall u \in \mathcal{U}_{v}.
\end{equation}
Since the dummy sub-task does not require any computing resource, it is the only sub-task being executed at the local. After the decision is made at the BS, the sub-tasks from the user have to be transferred to the executed UAV, and this process takes time. We define the latency to transmit the input data of sub-task $j$ as $t^\textrm{load}_{u,j}$. The arrival time of sub-task $j$ of user $u$ will be given by the following equation:  
\begin{equation}
    AT_{u,j}= {Trel_{u} + t^\textrm{{load}}_{u,j},\ \forall j \in N_{u}, \forall u \in \mathcal{U}_{v}. }
\end{equation}
The start time of a sub-task not only depends on the finish time of preceding sub-tasks $Pd_{u,j}$, the transmit latency of result data $D^{u}_{p,j}$ but also the arrival time of sub-task $j$. The relationship between the start time of the sub-task and those time variables can be expressed by the following equation:
\begin{equation}
\begin{split}
    ST_{u,j} \geq \max\bigg\{\max_{\forall p \in \textit{Pd}_{u,j}} \big(&FT_{u,p} + t^\textrm{{dep}}_{u,p,j}\big)\ ;\ AT_{u,j}\ \bigg\} \\
    &\forall p \in Pd_{u,j}, \forall j \in N_{u}, \forall u \in \mathcal{U}_{v}.
\end{split}
\end{equation}
The second way to include the transmitting time of the sub-task into our problem is to distribute the sub-task right after the decision is made at the BS. In this case, the start time can be given by:
\begin{equation}
\begin{split}
    ST_{u,j} \geq \max_{\forall p \in \textit{Pd}_{u,j}} \big(FT_{u,p} + &t^\textrm{dep}_{u,p,j}\big)\ + AT_{u,j}\\
    &\forall p \in Pd_{u,j}, \forall j \in N_{u}, \forall u \in \mathcal{U}_{v},
\end{split}
\end{equation}
where $t^\textrm{dep}_{u,p,j}$ is the latency for transmitting dependent data from other devices, which is the result of the preceding sub-task. In case sub-tasks $j$ needs multiple result data, we will have to wait until the last one is finished and transmitted to the UAV that is responsible for executing sub-task $j$. However, when all dependent sub-tasks of sub-task $j$ are located at the same device with sub-tasks $j$, this latency becomes zero.
The transmission latency for transmitting $D^{u}_{p,j}$ depends on the amount of data we need to send and the channel quality between UAVs, and it can be determined as: 
\begin{equation}
      t^{\textrm{dep}}_{u,p,j}= \frac{D^{u}_{p,j}}{R_{v,w}}, \ \forall p \in Pd_{u,j}, \forall j \in N_{u}, \forall u \in \mathcal{U}_{v},
\end{equation}
where $D^{u}_{p,j}$ is the dependent data between sub-tasks $p$ and $j$ of task $u$. $R_{r,w}$ is the achievable data rate between UAVs. 
Then, the finishing time for sub-task $j$ will be calculated by taking the sum of the start time and the execution time of it:\\
\begin{equation}
    FT_{u,j}= ST_{u,j} + t^\textrm{exe}_{u,j},\ \forall j \in N_{u}, \forall u \in \mathcal{U}_{v},
\end{equation}
where ${t^\textrm{exe}_{u,j}}$ will be the execution time of , which will be defined in much more detail later.
% \begin{equation}
%     % t^\textrm{exe}_{u,j}= t^{loc}_{u,j}_{u,j} + t^{v,\textrm{exe}}_{u,j} \theta^{u\rightarrow v}_{u,j} +t^{ w,\textrm{exe}}_{u,j}\gamma^{v\rightarrow w}_{u,j} + t^{bs,\textrm{exe}}_{u,j}\phi^{bs}_{u,j}
%     t^\textrm{exe}_{u,j}= \frac{C_{u} Z_{u,j}}{f^{v}_{u,j}} 
% \end{equation}

\subsubsection{Network Model} 
A number of sub-tasks will be offloaded to the associated UAV $v \in \mathcal{V}$ through the wireless communication link. According to \cite{zhou2017resource}, the achievable spectrum efficiency from user $u$ to UAV $v$ can be determined by the following equation:
\begin{equation}
    \Gamma^{u \rightarrow v}= \log_{2}\bigg(1+\frac{P_{u}G^{v}_{u}}{\sigma^{2}}\bigg),\ \forall u \in \mathcal{U}_{v},\forall v \in \mathcal{V},
\end{equation}
in which $P_{u}$ is the power of user $u$, $G^{v}_{u}$ is the channel gain between user $u$ and UAV $v$, while $\sigma^{2}$ is white the Gaussian noise power. Generally, the channel gain between mobile device $u$ and UAV $v$ is acquired by: 
\begin{equation}
    G^{v}_{u}= 10^{-{\delta^{v}_{u}}/{10}},\ \forall u \in \mathcal{U}_{v},\forall v \in \mathcal{V},
\end{equation}
where $\delta^{v}_{u}$ is the path loss between UAV and user $u$. Furthermore, in this work, we not only consider line of sight (LoS), but also non-line of sight (NLoS) links for the air-to-ground communication (in particular users to UAVs and backward). Then, the path loss $\delta^{v}_{u}$ is the combination of 2 types of path losses: LoS path loss, $\delta^{v,\textrm{LoS}}_{u}$ and non-LoS path loss, $\delta^{v,\textrm{NLoS}}_{u}$. The formulation of these path losses is defined in \cite{dinh2013survey} as below:
\begin{equation}
    \delta^{v,\textrm{LoS}}_{u}=2n  \log_{2}\bigg(\frac{4\pi d^{v}_{u} B^{\textrm{lte}}_{c}}{c}\bigg)+ \textit{L}_{\textrm{LoS}},
\end{equation}

\begin{equation}
    \delta^{v,\textrm{NLoS}}_{u}=2n  \log_{2}\bigg(\frac{4\pi d^{v}_{u} B^{\textrm{lte}}_{c}}{c}\bigg)+ \textit{L}_{\textrm{NLoS}},
\end{equation}
where n $\geq$ 2 is the path loss exponent, $B^{\textrm{lte}}_{c}$ denotes the carrier frequency (i.e., 2 GHz), c is the speed of light, ${L}_{\textrm{LoS}}$ and ${L}_{\textrm{NLoS}}$ are the average added losses for LoS and NLoS links. Additional, $d^{u,v}$ is the distance from mobile user to UAV, and can be acquired by the following equation: 

\begin{equation}
    d^{u,v}= \sqrt{(x_{v}-x_{u})^2+(y_{v}-y_{y})^2+z^2_{v}}.
\end{equation}
The next thing we consider is the probability of existing LoS connectivity between UAV $v$ and user $u$, which is denoted in \cite{mozaffari2016unmanned} as follows: 
\begin{equation}
    Pr^{v,\textrm{LoS}}_{u}= \frac{1}{1 + C.exp\bigg[-D(\frac{180}{\pi}tan^{-1}(\frac{z_{v}}{d^{u,v}})-C)\bigg]},
\end{equation}
where $z_{v}$ is the hovering altitude of UAV $v$. While $C$ and $D$ are variables that depend on the operating environment, such as city or countryside and others. As a result, the likelihood of the existing NLoS link from mobile user to the associated UAV is acquired by the following equation: 
\begin{equation}
    Pr^{v,\textrm{NLoS}}_{u}= 1 - Pr^{v,\textrm{LoS}}_{u}.
\end{equation}
Then, the path loss from user $u$ to the associated UAV is calculated by the following equation: 
\begin{equation}
     \delta^{v}_{u}= Pr^{v,\textrm{LoS}}_{u}\delta^{v,\textrm{LoS}}_{u} +  Pr^{v,\textrm{NLoS}}_{u} \delta^{v,\textrm{NLoS}}_{u} ,\forall u \in \mathcal{U}_{v},\forall v \in \mathcal{V}.
\end{equation}
Back to the spectrum efficiency between user $u$ and UAV $v$, the achievable uplink transmission rate of device $u$ will be determined as below:
\begin{equation}
     R^{u\rightarrow v}=\beta^{v}_{u}B^{v}\Gamma^{u \rightarrow v},\forall u \in \mathcal{U}_{v},\forall v \in \mathcal{V},
\end{equation}
where $\beta^{v}_{u}$ is the percentage of total available bandwidth of UAV $v$ (denoted as $B^{v}$) that will be allocated to user $u$.

Thus, the transmission latency from user $u$ to UAV $v$ depends on uplink transmission rate and can be formulated as follows:
\begin{equation}
    t^{u\rightarrow v}=\frac{ \sum_{j=1}^{\mathcal{T}_{u}} H_{u,j}}{R^{u\rightarrow v}},
\end{equation}
where $H_{u,j}$ is the input size of sub-task $j$ belong to task $u$. Then the energy consumed by up-link transmitting from user  to UAV is denoted as follows:
\begin{equation}
    E^{v,\textrm{up}}_{u}=\frac{P_{u}\sum_{j=1}^{\mathcal{T}_{u}} H_{u,j}}{R^{u \rightarrow v}}.
\end{equation}
The offloaded sub-tasks are executed on which UAVs are decided at the BS. The decision variable representing that the sub-task will be offloaded to the associated UAV $v$ is defined as follow: 
\begin{equation}
    \theta^{v}_{u,j} =\left \{ \begin{array}{ll}{1,} & {\textrm{if sub-task $j$ of user $u$ is offloaded to }}  \\ 
     & {\textrm{ associated UAV $v$,}} \\{0,} & {\textrm{otherwise.}}\end{array}\right.
\end{equation}
In this case, the latency for transmitting the input data of sub-task $j$ will given as bellow:
\begin{equation}
    t^{\textrm{load}}_{u,j}= t^{u\rightarrow v}_{u,j}.
\end{equation}
When the sub-task is executed at UAV $v$, the computing time of that sub-task can be given as:
\begin{equation}
     t^{{\textrm{exe}}}_{u,j} = \frac{C_{u}H_{u,j}}{f^{v}_{u,j}}, \forall u \in \mathcal{U}_{v},\forall v \in \mathcal{V},
\end{equation}
where $f^{v}_{u,j}$ is the computing resource (i.e., cycles/s) of the MEC server mounted on UAV $v$ that is allocated to execute sub-task $j$ of user $u$. This computing resource will be determined based input size of sub-task $j$ over all the sub-task executed at UAV $v$ as defined in \cite{tun2019wireless}:
\begin{equation}
    f^{v}_{u,j}=\frac{H_{u,j}}{\sum_{q\in U_{v}} H_{u,j}} F^{\textrm{max}}_{v},
\end{equation}
where $F^\textrm{max}_{v}$ is the maximum computing resource available of UAV $v$. We have to ensure that the total computing capacity of UAVs across number of sub-tasks from different users is less than or at least equal to the maximum capacity of the UAV, i.e.,
\begin{equation}
    \sum_{q\in U_{v}} \theta^{v}_{u,j} f^{v}_{u,j} \leq F^{\textrm{max}}_{v}, \forall v \in \mathcal{V}.
\end{equation}
% In summary, the total latency to finish the sub-tasks when they are offloaded to the associated UAV is given by this equation: 
% \begin{equation}
%     t^{v,e}_{u,j}=t^{v,up}_{u,j}+t^{v,\textrm{exe}}_{u,j}+t^{dep}_{u,jv}
% \end{equation}
% where $t^{dep}_{u,jv}$ is the latency for transmit dependent data from preceding sub-tasks. This latency depends on the location in which preceding sub-tasks are allocated. Therefore, it can be formulated by the following equation:\textcolor{red}{(fixed this)  Multiple dependent times} 
% \begin{equation}
%     t^{dep}_{u,jv}=\max\bigg[\frac{D^{u}_{p,j}}{R^{u \rightarrow v}} _{up};\frac{D^{u}_{p,j}}{R^{w \rightarrow v}}_{up} \gamma^{v \rightarrow w}_{up};\frac{D^{u}_{p,j}}{R^{bs \rightarrow v}}\phi^{v \rightarrow bs}_{up}\bigg]
% \end{equation}
% where these fractions are dependent latency from user $u$, UAV $w$ and base station to UAV $v$, respectively. We will need to wait until received that last dependent results from all the sources. 
% The next thing is the energy consumption needed to execute the offloaded sub-task of user v is given by:

The following equation will be used to calculate how much energy of UAV $v$ will use to complete sub-task $j$, i.e., 
\begin{equation}
    E^{v,\textrm{exe}}_{u,j}=\texttt{k}_{v} (f^{v}_{u,j})^{2}C_{u}Z_{u,j},\forall u \in \mathcal{U}_{v},
\end{equation}
where $\texttt{k}_{v}$ = 5 $\times$ $10^{-27}$ is a constant that only changes according to the server's chip architecture mounted on the UAV.

UAV $v$ will pass on a part of the task or all of them to other UAVs ($w \in \mathcal{V}$, $v \neq w$) in the network by wireless link if its computing capacity is not sufficient to execute the task alone.
% When the  computing capacity at UAV $v$ is not sufficicent in order to compute the offloaded sub-tasks of its associated users, the UAV $v$ forwards the tasks to the neighboring UAVs, $w \in $v$$, $v \neq W$ via wireless.
Therefore, we need to define another binary variable $\gamma^{w}_{u,j}$ as a decision variable, which meaning is whether or not to forwards the computing sub-tasks of user $u$ from UAV $v$ to UAV $w$:
\begin{equation}
    \gamma^{w}_{u,j} =\left \{ \begin{array}{ll}{1,} & {\textrm{if sub-task $j$ of user $u$ is forwarded to }}  \\ 
     & {\textrm{UAV $w$,}} \\{0,} & {\textrm{otherwise.}}\end{array}\right.
\end{equation}
The latency for the associated UAV $v$ to transmit sub-task $j$ to other UAV is given as follows:
\begin{equation}
    t^{v\rightarrow w}_{u,j}= \frac{H_{u,j}}{R^{v\rightarrow w }},
\end{equation}
where $R^{v\rightarrow w}$ is the achievable rate from UAV $v$ to UAV $w$ defined as:
\begin{equation}
    R^{v \rightarrow w}= B^{v,w} \log_{2}{\bigg(1+\frac{P_{v}G_{v ,w}}{{\sigma}^2}\bigg)},
\end{equation}
where $B^{v, w}$ is the available bandwidth for direct communication between UAVs $v$ and $w$, $P_{v}$ is the power that UAV $v$ use for transmission, and $G^{v,w}$ is the achievable channel gain between two UAVs. Because the UAVs fly over the sky and therefore we assume line-of-sight (LoS) link for UAV-to-UAV communication. Then the channel gain between UAVs, which is defined in \cite{challita2017network}, as follows:
\begin{equation}
    G^{v,w}= 10^{-(L_{v,w}/10)},
\end{equation}
where $L_{v,w}$ = $\Theta_{v, w} + \Gamma_{LoS}$ is the pathloss between UAVs $v$ and $w$. Here, $\Gamma_{LoS}$ is the additional attenuation factor for LoS link \cite{challita2017network}, and $\Theta_{v, w}$ is as follows: 
\begin{equation}
    \Theta_{v,w}(dB)= 20 \log_{10}({d^{v,w}}) + 20\log_{10}({f_{c}}) + 10 \log_{10}\bigg[\bigg({\frac{2\pi}{c}}\bigg)^2\bigg]\\,
\end{equation}
where c is the speed of light, $f_{c}$ is the carrier frequency (i.e., 2 GHz). Furthermore, $d^{w}_{v}$ is the distance between UAVs $v$ and $w$, and it can be expressed as follows:
\begin{equation}
    d^{v,w}= \sqrt{(x_{w}-x_{v})^2+(y_{w}-y_{v})^2+(z_{w}-z_{v})^2},
\end{equation}
where $[x_{w},y_{w},z_{w}]$ is coordinate vector of UAV $w$ in three-dimensional area.\\
In case sub-task $j$ belongs to user $u$ being offloaded to the collaborative UAV $w$, the latency for transmitting input data will be the sum over two components: transfer time from user to associated UAV $v$ and forward time to the collaborative UAV $w$. Therefore, it can be given by: 
\begin{equation}
    t^\textrm{load}_{u,j}= t^{u\rightarrow v}_{u,j} + t^{v\rightarrow w}_{u,j}.
\end{equation}
The energy consumed for transmitting input data from UAVs $v$ to $w$ in the network is given by:
\begin{equation}
    E^{v \rightarrow w}_{u,j}= P_{v} \bigg( \frac{H_{u,j}}{R^{v\rightarrow w }}  \bigg) , \forall u \in \mathcal{U}_{v},\forall v \in \mathcal{V}.
\end{equation}
Once UAV $w$ received sub-task $j$ of user $u$, the latency and energy consumption, $t^{\textrm{exe}}_{u,j}$ and $E^{w,\textrm{exe}}_{u,j}$, when sub-task $j$ is executed here can be easily calculated based on (22) and (25).\\
The total energy usage of UAV $v$ will be the sum of the computation energy to execute sub-tasks, the energy to forward the input data of sub-tasks to neighboring UAVs, the energy to transmit information to BS, and finally the energy consumption to hover at a fixed altitude. Furthermore, the length of the input data is much longer than the dependent data, so that we omit the energy for transmitting the dependent data. The total energy consumption can be presented as: 
\begin{equation}
    E^\textrm{tol}_{v}= \sum_{u\in U_{v}}{\theta^{u}_{u,j}}E^{v,\textrm{exe}}_{u,j} + \sum_{v\in W, w\neq v}\gamma^{w}_{u,j}E^{v\rightarrow w} + E^{v\rightarrow b} + E^{v,\textrm{hov}},
\end{equation}
where $E^{v,\textrm{hov}}$ is the hovering energy of UAV $v$ and calculated as follows \cite{monwar2018optimized}:
\begin{equation}
    \begin{aligned}
    E^{v,\textrm{hov}} &= P^{v,\textrm{hov}}t^{v,\textrm{hov}}\\
    &=\frac{\eta\sqrt{\eta}}{\varphi_{v}\sqrt{2\pi q r^{2} \varkappa}} t^{v,\textrm{hov}},
    \end{aligned}
\end{equation}
where $\eta$ is the trust that is proportional to the UAV's mass, $\varphi_{v}$ is the power efficiency of UAV $v$, $q$ denotes the number of rotors belong to a single UAV, $r$ is the diameter of each rotor, and $\varkappa$ is the air density. Finally, $t^{v,\textrm{hov}}$ is the maximum hovering time of the UAV, and it is given by:
\begin{equation}
\begin{split}
    t^{v,\textrm{hov}}=\max_{u\in U_{v}}\bigg[ t^{v,\textrm{up}}_{u}+ t^{v\rightarrow b} +\\ \max\bigg(&t^{v,\textrm{exe}}_{u};t^{v\rightarrow w,\textrm{com}}_{u}+t^{v \rightarrow w,\textrm{exe}}_{u}\bigg) \bigg].
\end{split}
\end{equation}

As mentioned before, the BS operates as a centralized controller and is responsible for offloading decisions. To do this, the MEC-enabled UAVs have to send their information: maximum computing capacity, available energy, along with the directed acyclic graph of the tasks, which is denoted as $I_{v}$. The communication latency from UAV $v$ to the ground BS for transmitting that information can be acquired by: 
\begin{equation}
    t^{v\rightarrow b}=\frac{I_{v}}{R^{v\rightarrow b}}, \forall v \in \mathcal{V},
\end{equation} 
where $R^{v\rightarrow b}$ is the achievable data rate of the mmWave link between the ground BS and UAV $v$, which can be calculated by \cite{lai2019data}:
\begin{equation}
    R^{v\rightarrow b}= B^{v\rightarrow b} \log_{2}\left(1+\frac{P_{b,v}}{B^{v\rightarrow b}_{\textrm{mm}}\sigma^{2}}\right), \forall v \in \mathcal{V},
\end{equation}
where $B^{v\rightarrow b}_{\textrm{mm}}$ is the bandwidth in mmWave frequency band that the BS allocated to UAV $v$. $P_{b,v}$ is the received power at the BS, and calculated as follows: 
\begin{equation}
    P_{b,v}=P^{v\rightarrow b} G^{tx}_{v} G^{rx}_{b} \frac{c}{4\pi d^{b}_{v} B^{\textrm{mm}}_{c}},
\end{equation}
where $P^{v\rightarrow b}$ is transmit power to the BS and $B^{\textrm{mm}}_{c}$ is the carrier frequency of the mmWave link. The distance $d^{b}_{v}$ from UAV $v$ to the BS is determined by: 
\begin{equation}
    d^{b}_{v}= \sqrt{(x_{b}-x_{v})^2+(y_{b}-y_{v})^2+(z_{v})^2},
\end{equation}
where $[x_{b},y_{b}]$ is the coordinate of the BS. From a couple of previous denotations, the energy consumption to transmit information from UAV $v$ to the BS is defined as follows: 
\begin{equation}
    E^{v\rightarrow b}= P^{v\rightarrow b}\frac{I_{v}}{R^{v\rightarrow b}}, \forall v \in \mathcal{V}.
\end{equation}

The constraint which controls each sub-task is offloaded to only one UAV in the network can be described by:
\begin{equation}
     \theta^{v}_{u,j} + \gamma^{w}_{u,j}  = 1, \forall j \in N'_{u}, \forall u \in \mathcal{U}_{v}, \forall w  \in \mathcal{V}.
\end{equation}

The completion time of a task will be represented by (43), which is the difference between the time instance when the last sub-task is finished and the time instance when the task is released:
\begin{equation}
    F_{u}= \max_{j=1,...,N_{u}} FT_{u,j}- ST_{uN^{'}}.
\end{equation}

To make the problem formulation easy to follow, we denote one more variable, which is the ready time of a sub-task:
\begin{equation}
    RT_{u,j}=\max\bigg\{\max_{\forall p \in \textit{Pd}_{u,j}} \big(FT_{u,p} + t^\textrm{{dep}}_{u,p,j}\big)\ ;\ AT_{u,j}\ \bigg\}.
\end{equation}

% \textcolor{red}{Not yet include power transmit for dependent data. However we can say that compare to the power transmit for input data, the dependent data can be much more smaller, so we can omit that energy in our problem}
\subsection{Problem Formulation}
In this paper, we investigate a realistic joint offloading decision and communication resource allocation problem with the objective of minimizing the average latency of all users in the network. Therefore, the objective function will be defined as $M(\theta,\gamma,\beta)$ = $\frac{1}{U_{v}} \sum\limits_{u \in U_{v}} \bigg\{ F_{u}(\theta,\gamma) +t^{u\rightarrow v}(\beta)\bigg\}$. At the time, we are working on this paper, there is no paper that both consider solving the offloading decision for dependency task and wireless resource allocation (communication and computing) in a network with multiple MEC-enabled UAVs. The problem can be mathematically presented as:
\begin{mini!}|l|
	{\boldsymbol{\theta},\boldsymbol{\gamma},\boldsymbol{\beta}}{M(\theta,\gamma,\beta)}
	{}{}
	\addConstraint{\theta^{v}_{u,j} + \gamma^{w}_{u,j}  = 1}\nonumber
	\addConstraint{ \qquad\qquad \forall j \in N'_{u}, \forall u \in \mathcal{U}_{v}, \forall w  \in \mathcal{V}}\label{c1}
	\addConstraint{\theta^{v}_{u,j}\in\{0,1\},\forall j \in N'_{u}, \forall u \in \mathcal{U}_{v}}\label{c2}
	\addConstraint{\gamma^{w}_{u,j}\in\{0,1\},\forall j \in N'_{u}, \forall u \in \mathcal{U}_{v}, \forall w  \in \mathcal{V}}\label{c3}
	\addConstraint{\sum_{u\in U_{v}} \beta^{v\rightarrow w} \leq 1,\forall{v} \in \mathcal{V}},\forall{u} \in \mathcal{U}\label{c4}
	\addConstraint{\beta^{v\rightarrow w} \in [ 0,1 ] , \forall u \in \mathcal{U}_{v} }\label{c5}
	\addConstraint{E^\textrm{tol}_{v} \leq E_v^{\textrm{max}},  \forall v \in \mathcal{V},}\label{c6}
% 	\addConstraint{ST_{u,j} \geq  RT_{u,j} }
\end{mini!}
where $\theta$ is the offloading decision vector which has $\sum\limits_{u=1}^{K} \sum\limits_{j=1}^{N_{u}}1$ elements and each element represents whether sub-task $j$ of task $u$ offload to associated UAV or not, while $\gamma$ is offloading decision vector to collaborative UAVs in our network and the size of this vector will be $\sum\limits_{w \in \mathcal{V}, w\neq v} \sum\limits_{u=1}^{K} \sum\limits_{j=1}^{N_{u}}1$, $\beta$ is the communication resource (i.e., bandwidth) allocation vector with each element $\beta^{v}_{u}$ represents the fraction of bandwidth allocated to user $u \in \mathcal{U}_{v}$ at UAV $v \in \mathcal{V}$. The constraint (45b) ensures the sub-task has to be executed at one device and constraint (45c) and (45d) indicate the binary decisions variables. 
constraint (45e) and (45f) guarantee that the total fraction bandwidth allocated to users should be less than the maximum available bandwidth at a UAV. Finally, constraint (45g) represents the energy constraint of each UAV.

In the problem, we have to assign each sub-task of users' tasks to one MEC-enabled UAV, which can be simplified as a Generalized Assignment Problem (GAP). GAP is a NP-hard problem and therefore, our problem is also a NP-hard problem. We cannot find the optimal solution in polynomial time. Therefore a heuristic solution will be the best way to solve the problem.

\section{proposed solution}\label{Proposed}
In order to address the formulated NP-hard problem, we first decompose the problem into two sub-problems by adopting the block coordinate descent technique: 1) offloading dependency task problem and 2) communication resource allocation problem. Then, a meta-heuristic approach so-called discrete whale optimization algorithm to solve sub-problem \ref{Offloading} and standard optimization tool is applied to solve sub-problem \ref{Allocate}, respectively. 
% By adopting the block coordinate descent (BCD) technique, the above problem is decomposed into 2 tractable sub-problems.
% By using the block coordinate descent (BCD) method, the aforementioned problem is divided into two tractable sub-problems.

% After that with each sub-problem, we will propose an appropriate solution.
\subsection{Offloading dependency task problem}\label{Offloading}
Given a fixed bandwidth allocation and start time of a sub-task, our first sub-problem to find offloading decisions will be formulated as follows:
\begin{mini!}|l|
	{\boldsymbol{\theta},\boldsymbol{\gamma}}{M(\theta,\gamma)}
	{}{}
	\addConstraint{\theta^{v}_{u,j} + \gamma^{w}_{u,j}  = 1}\nonumber
	\addConstraint{ \qquad\qquad \forall j \in N'_{u}, \forall u \in \mathcal{U}_{v}, \forall w  \in \mathcal{V}}\label{c7}
	\addConstraint{\theta^{v}_{u,j}\in\{0,1\},\forall j \in N'_{u}, \forall u \in \mathcal{U}_{v}}\label{c8}
	\addConstraint{\gamma^{w}_{u,j}\in\{0,1\},\forall j \in N'_{u}, \forall u \in \mathcal{U}_{v}, \forall w  \in \mathcal{V}}\label{c9}
	\addConstraint{ E^\textrm{tol}_{v} \leq E_v^{\textrm{max}},  \forall v \in \mathcal{V}.}\label{c10}\
% 	\addConstraint{ST_{u,j} \geq  RT_{u,j} }
\end{mini!}

By observing, we can see the coupling between $\theta$ and $\gamma$ in (46b), and beyond that, they are binary variables. With these statements, we can conclude the proposed problem is an NP-hard problem. Normally, we can relax these binary variables into continuous and then apply the convex optimization method to tackle the problem. However, with $\mathcal{N}$ sub-tasks in a single task, the number of offloading decisions will increase exponentially and become impossible to solve in polynomial time. Therefore, we use a meta-heuristic solution, whale optimization algorithm (WOA), to mimic humpback whales' social hunting behavior. Before deploying the algorithm, we need to merge the offloading variable $\theta$ and the offloading vector $\gamma$ into a single vector, $X_{u}=[x_{u1},x_{u2},...,x_{un}]$, where $X_{u}$ and $x_{uj}$ are the offloading vectors for task $u$ and offloading decision for sub-task $j$ of task $u$, respectively. The value of $x_{uj}$ is discrete and indicates which UAV is responsible for executing the sub-task. The reformulated problem is given as follows: 
\begin{mini!}|l|
	{\boldsymbol{\chi}}{M(\boldsymbol{X})}
	{}{}
	\addConstraint{ x_{uj} \in [ 1, V ], \forall j \in N'_{u},\forall u \in \mathcal{U}_{v}}\label{10}
% 	\addConstraint{ \qquad\qquad \forall j \in N'_{u},\forall u \in \mathcal{U}_{v}}\label{10}
	\addConstraint{ E^\textrm{tol}_{v} \leq E_v^{\textrm{max}},  \forall v \in \mathcal{V}}\label{11}.
% 	\addConstraint{ST_{u,j} \geq  RT_{u,j} }
\end{mini!}
\textbf{\textit{Whale Optimization Algorithm:}} In meta-heuristic algorithms, they are divided into four classes: Evolutionary algorithms, Physics-based algorithms, Swarm-based algorithms, and Human-based algorithms. WOA belongs to the swarm-based algorithm class, which has two phases in its searching process: \textit{exploration phase} and \textit{exploitation phase}. The former phase will help the optimizer explore the search space globally; and therefore, the movement of searching agents needs to be as randomized as possible in this phase. In contrast to this, the exploitation phase can be described as a process of examining the promising regions of the search space. 
The algorithm has been mathematically modeled based on the search for prey of humpback whales by \cite{MIRJALILI201651}. Humpback whales have the ability to detect the position of their preys, when they find out the prey, they swim around and surround them by creating a bubble in a spiral shape \cite{li2020novel}. The algorithm has three main mechanisms: encircling, bubble-net hunting, and finally, searching. 
\subsubsection{Encircling prey}
At the beginning of the search, the optimal solution is unknown to agents, and the algorithm considers the current best solution as the target prey so that other search agents will update their solutions toward the best searching agent. The following equations illustrate this behavior: 
\begin{subequations}
\begin{align}
\Vec{D}= \lvert \Vec{C} \cdot \Vec{X}^{*}(t)-\Vec{X}(t)\lvert, \\
\Vec{X}(t+1)=  \Vec{X}^{*}(t)-\Vec{A}\cdot \Vec{D},
\end{align}
\end{subequations}
where ${\Vec{X}^{*}}(t)$ denotes the current best solution, and it will be updated in each iteration if any agents find out better objective, $\Vec{C}$ and $\Vec{A}$ are co-efficient vectors, $t$ represents the number of the current iteration, the math operation $\cdot$ indicate element by element multiplication. Vectors $\Vec{A}$ and $\Vec{C}$ can be given by the following equations:
\begin{subequations}
\begin{gather}
\Vec{A}= 2 \Vec{a} \cdot \Vec{r}-\Vec{a}, \\
\Vec{C}=2\cdot\Vec{r},
\end{gather}
\end{subequations}
where $\Vec{a}$ is linearly decreased from 2 to 0 over the course of iterations and $\Vec{r}$ is a random vector in [0,1]. This mechanism is called shrinking encircling in the original whale optimization algorithm paper.
\subsubsection{Bubble-net hunting}
As mentioned before, whale agents can locate the position of prey and swim to the prey along a spiral path. This hunting mechanism can prevent prey from escaping and can be mathematically modeled as follows:
\begin{subequations}
\begin{gather}
    \Vec{X}(t+1)= \Vec{D}' \cdot e^{bl} \cdot cos (2\pi l) + \Vec{X}^{*}(t),\\
\Vec{D}'= \lvert \Vec{X}^{*}(t)-\Vec{X}(t) \lvert,
\end{gather}
\end{subequations}
where $\Vec{D}'$ is the distance between the \textit{i}th whale and the optimal solution obtained so far, \textit{b} is the constant for defining the shape of the logarithmic spiral, and \textit{l} is a random number between [-1,1].

The humpback whales not only swim around the prey within a shrinking circle but also along a spiral-shaped path at the same time. In order to capture this simultaneous behavior, a random variable \textit{p} is introduced. Its value falls in [0,1], and we will choose an action based on its value. The mathematical model is given as follows:
\begin{equation}
    \Vec{X} (t+1)=\left \{ \begin{array}{ll}{\Vec{X}^{*}(t)-\Vec{A}\cdot\Vec{D},} & {\textrm{p $<$0.5,}}  \\ 
      \\{\Vec{D}' \cdot e^{bl} \cdot cos (2\pi l) + \Vec{X}^{*}(t),} & {\textrm{p $\geq$0.5.}}\end{array}\right.
\end{equation}
The above equation will give both behaviors an equal probability of being picked. 
\subsubsection{Searching for prey} The approach for the shrinking encircling mechanism will be re-used for the search (exploration stage). When agents are in this stage, they search randomly according to the position of each other. The mechanism will only be active when $\lvert \Vec{A} \lvert \geq$ 1, the search agent updates its position far away from the reference whale, which is the difference from the encircling mechanism when $\lvert \Vec{A} \lvert < $ 1. The mathematical model of the searching phase will be as follows:
\begin{subequations}
\begin{align}
\Vec{D}= \lvert \Vec{C}\cdot\Vec{X}_{rand}(t)-\Vec{X}(t)\lvert, \\
\Vec{X}(t+1)=  \Vec{X}_{rand}(t)-\Vec{A}\cdot\Vec{D}.
\end{align}
\end{subequations}
\subsubsection{Objective function}

With each decision matrix that we obtain from the searching agents, we calculate the total latency, which will be the sum of distributed latency and computing latency. The distributed latency is the time to transmit the sub-tasks from the associated MEC-enabled UAV to the other MEC-enabled UAVs in the network. However, there is one important detail is that not all of the solution given by searching agents is feasible due to energy constraint, while the original WOA is designed to address the unconstrained optimization problem. Therefore, we have to deal with the constraints, in \cite{pham2020whale} the authors introduce some constraint-handling techniques such as the penalty method, feasibility rules, stochastic ranking and split them into categories. In this paper, we adopt the most simple and also well-known technique: the penalty method to our objective. Therefore, the reformulated problem in (42) is transformed into penalty form as follows:
\begin{mini!}|l|
	{\boldsymbol{\chi}}{M(\boldsymbol{X})+\lambda \cdot G(E^\textrm{tol}_{v}, E_v^{\mathbf{max}})\cdot(E^\textrm{tol}_{v}- E_v^{\mathbf{max}})^2}
	{}{}
	\addConstraint{ x_{uj} \in [ 1, V ], \forall j \in N'_{u}, \forall u \in \mathcal{U}_{v},}
\end{mini!}
where G($E^\textrm{tol}$, $E^{\mathbf{max}}$) is a step function condition by $E^\textrm{tol}$ and $E^{\mathbf{max}}$, if the energy consumption given by offloading decision exceeds the maximum energy of any UAV in the network, the function G($\cdot$) returns the value 1, otherwise it equals to 0. The right term in our objective function is denoted as the penalty term, and $\lambda$ is the penalty coefficient, which is normally adopted to manipulate the penalty value. If the penalty's value is too small, then the proposed algorithm may converge to an infeasible solution and accept the punishment. On the other hand, the severe penalty prevents the agent from exploring more promising regions and staying in the comfort zone. 
\newfloat{algorithm}{t}{lop}
\begin{algorithm}
\caption{Discrete Whale Optimization Algorithm for Offloading Decisions}\label{alg:cap}
\hspace*{\algorithmicindent} \textbf{Input} \textit{N}: Number of searching agents, $v$: Number of UAVs in the network, \textit{n} total number of sub-tasks and \textit{MaxIT}.
\hspace*{\algorithmicindent} \textbf{Output} Optimal decision Y* and the objective function value \textit{f}(Y*).
\begin{algorithmic}[1]
\State Initialize and randomly generate: 
\newline $X_{u}$ = [$x_{u1}, x_{u2}, ..., x_{un}$] $\in$ $[1,V]^{n}$ $(u=1,2,...,N)$ , 
\newline Calculate latency of each decision, determine X*, $t \gets 0$.
\While{$t \leq MaxIT $}
    \For{$u \gets 1 $ to \textit{N}}
    \State Update vector \textit{a, A, C, l} and \textit{p}.
    \If{$p < 0.5$}
        \If{$ |A| < 1$}
            \State Update the offloading decision by (48b).
        \Else
            \State Update the offloading decision by (52b).
        \EndIf
    \ElsIf{$p\geq 0.5$}
    \State Update position by (49a).
    \EndIf
    \EndFor
    \State Calculate the latency of each decision matrix and update X*.
    \State $t \gets t +1$. 
\EndWhile
\State Return(X*, f(X*)).
\end{algorithmic}
\end{algorithm}

\subsection{Communication Resource Allocation}\label{Allocate}
In the beginning, users only send their task information to the associated UAV using a limited sub-channel. However, when UAVs have all the knowledge of active users in the coverage area, we can re-allocate the bandwidth between UAVs and their associated users by taking advantage of idling bandwidth to minimize the communication latency between users and UAVs:
\begin{mini!}|l|
	{\boldsymbol{\beta}}{M(\beta)}
	{}{}
	\addConstraint{\sum_{u\in U_{v}} \beta^{v\rightarrow u} \leq 1,\forall{v} \in \mathcal{V}}\label{c11}
	\addConstraint{\beta^{v\rightarrow u} \in [ 0,1 ] , \forall u \in \mathcal{U}_{v}, \forall v \in \mathcal{V}.}\label{c12}
\end{mini!}
With a fixed offloading decision matrix, the optimization problem in (54) becomes a convex problem. Therefore, we solve the problem by using the CVXPY toolkit.
\begin{algorithm}
\caption{The sequence of proposed scheme}\label{alg:cap1}
\begin{algorithmic}
\State \textbf{Step 1} Tasks are generated at the users site, the BS collect task information $D_{u}$ by using UAV as relay station.
\State  \textbf{Step 2}
\State  BS run D-WOA to acquire the optimal solution for offloading problem.
\State To reserve energy of UAVs, the optimal resource allocation from UAV to users is also solved by the BS. 
\State  \textbf{Step 3}
\State Users transmit the tasks to associated UAV through optimal bandwidth resource and then distributed to other UAVs in the network based on offloading decision from step 2. 
\end{algorithmic}
\end{algorithm}
\subsection{Complexity Analysis of the proposed algorithm}
The detail of the proposed scheme is given in Algorithm 2. In \cite{aung2022energy} and \cite{pham2020whale}, the authors state that the solution to each sub-problem can be used to determine the complexity of the proposed algorithm. Here we have two sub-problems: offloading decisions and communication resource allocation problems. For the first sub-problem, the computation complexity of the D-WOA normally depends on the number of searching agents $N$, the number of iterations $MaxIT$, and finally, the dimension of the searching agents, particularly in our case, is the number of sub-tasks $\mathcal{O}_{1}(N \cdot MaxIT \cdot M)$. However, due to the power constraint, the complexity of D-WOA increases and becomes $\mathcal{O}_{1} (N\cdot MaxIT\cdot(M+m))$, where $m$ is the number of inequality constraints. In the resource allocation problem, the complexity of it is quite simple $\mathcal{O}_{2} (V\cdot U_{v})$. Therefore, the computation complexity of our proposed approach to solve the problem is $\mathcal{O} (N MaxIT\cdot (M+m)+V\cdot U_{v})$.

\section{Performance Evaluation}\label{Performance}
In this section, we conduct a series of simulations to evaluate the performance of our algorithm and compare it with other benchmark solutions.

\subsection{Simulation Settings}
\subsubsection{Settings for Network Model}
We consider a rectangular region with the size of 1 km $\times$ 1 km, and it will be divided into four non-overlapping areas, each of them will be covered by a UAV at an altitude of around $50$ m. The number of users in each subarea being served by the MEC-enabled UAV will be randomly selected from a uniform distribution [2, 10] users, and notice that not all of them have a task to execute at the same time. The BS is located at the center of the area and within the communication range of all the UAVs. The detail of simulation parameters will be provided in Table  \ref{Simulationparameter}.

\begin{table}[t]
	\centering
	\caption{Simulation parameters}
	\begin{tabular}{|l|c|l|c|}
		\hline
		\textbf{Parameter} & \textbf{Value}&  \textbf{Parameter} & \textbf{Value}\\\hline
		$B^{v}$  & 3 MHz& $B^{v,w}$		& $8$ MHz \\ \hline 
		$P_{v}$  & 30 dBm &	$P_{u}$ & 23 dBm  \\ \hline
	    $f_{c}$	& $2$ GHz & $ F^\textrm{max}_{v}$ 	&  [800, 1000] MHz \\ \hline 
		$\sigma^{2}$	 & -174dBm &$ N $		 & 100 agents 	 \\ \hline
		$ MaxIT $ & 50 & 	$\eta$ & 30 N  	\\ \hline
		$j$	 & 	4 &	$r$ & 0.254   \\\hline
		$\varphi_{v}$ & 70 \% &	$D$		 & 0.1	\\ \hline
		$C$ & 11.9 &	$V$		 & 4	\\ \hline
	\end{tabular}
	\label{Simulationparameter}
\end{table}

\subsubsection{Settings for Task Model}
The topology of the task will be generated layer-by-layer, and the number of sub-tasks in each layer follows a normal distribution with $\mu=2$ and $\sigma^{2}=1$. Instead of randomizing the input size of the whole task, we control the size of sub-tasks, which are selected from Gaussian distribution with a mean value of 6 MB (Mega Byte), and the variance value is 1 MB. The dependence information among sub-tasks is drawn from a uniform distribution [150, 250] Kb. Finally, for the default case, each task has 10 sub-tasks, and the number of active users (have a task for offloading) is 3. 
\subsection{Benchmark Solutions}
To assess the performance of our proposal, we compare our solution approach with the following schemes in terms of total latency:
\begin{itemize}
\item \textit{Associated UAV}: As the name it sounds, in this solution approach, the task generated by users will be transmitted to the associated UAVs of those users. It should be pointed out that one user is only associated with one UAV, and there is no collaboration between MEC-enabled UAVs in this case. Therefore, each MEC-enabled UAV has to execute the tasks from the associated users by itself.
\item  \textit{Exhaustive Search Algorithm}: The offloading problem in (41) can be considered as a knapsack problem, however, instead of a single knapsack we have a set of knapsacks (UAVs). A sub-task only needs to be executed by one UAV, thus the offloading decision matrix is restricted by this and also the energy constraint. We attempt to generate all feasible solutions, and among them find the solution that gives us the lowest latency.
\end{itemize}

\begin{figure}
    \centering
    \includegraphics[width=.52\textwidth]{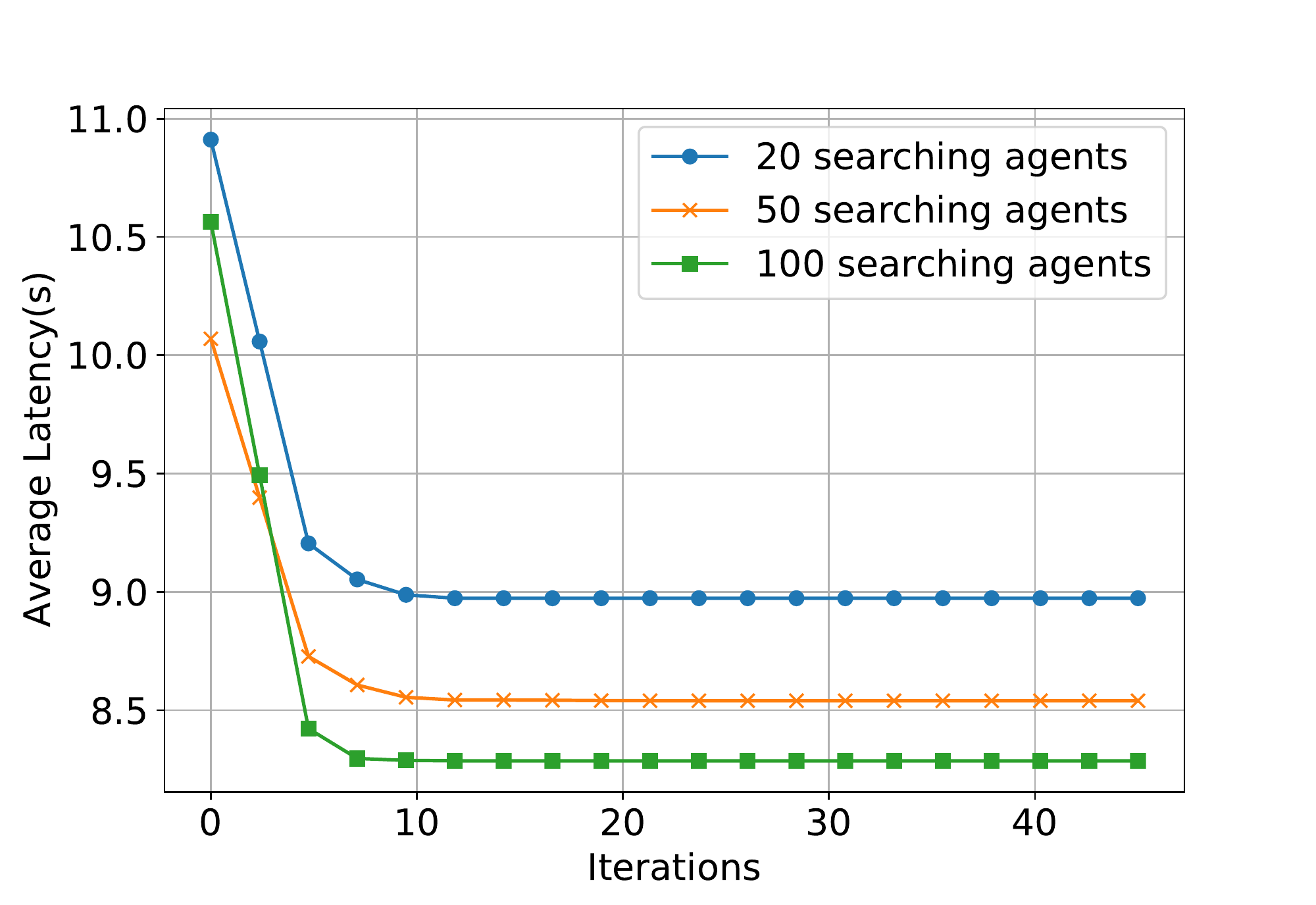}\
    \caption{Convergence of the proposed Whale optimization with different numbers of searching agents.}
    \label{fig:Result1}
\end{figure}

Fig.~\ref{fig:Result1} illustrates the convergence of the D-WOA under the different numbers of searching agents. As mentioned before, the heuristic solution can solve the NP-hard problem in polynomial time, but the solution is just local optimal. As shown in Fig.~\ref{fig:Result1}, when the number of searching agents increases, the optimal value also increases (i.e., achieve lower average latency when we increase the number of searching agents). This phenomenon can be easily understood that the more agents participate in the search, the more likely they encounter a better solution. This average latency is calculated by using the offloading decision belonging to the best searching agent. In addition, most of the improvements are made in the early iterations, in which the value of $\lvert \Vec{A} \lvert $ is large, indicating the agents are in the exploration stage. 
\begin{figure}
    \centering
    \includegraphics[width=.48\textwidth]{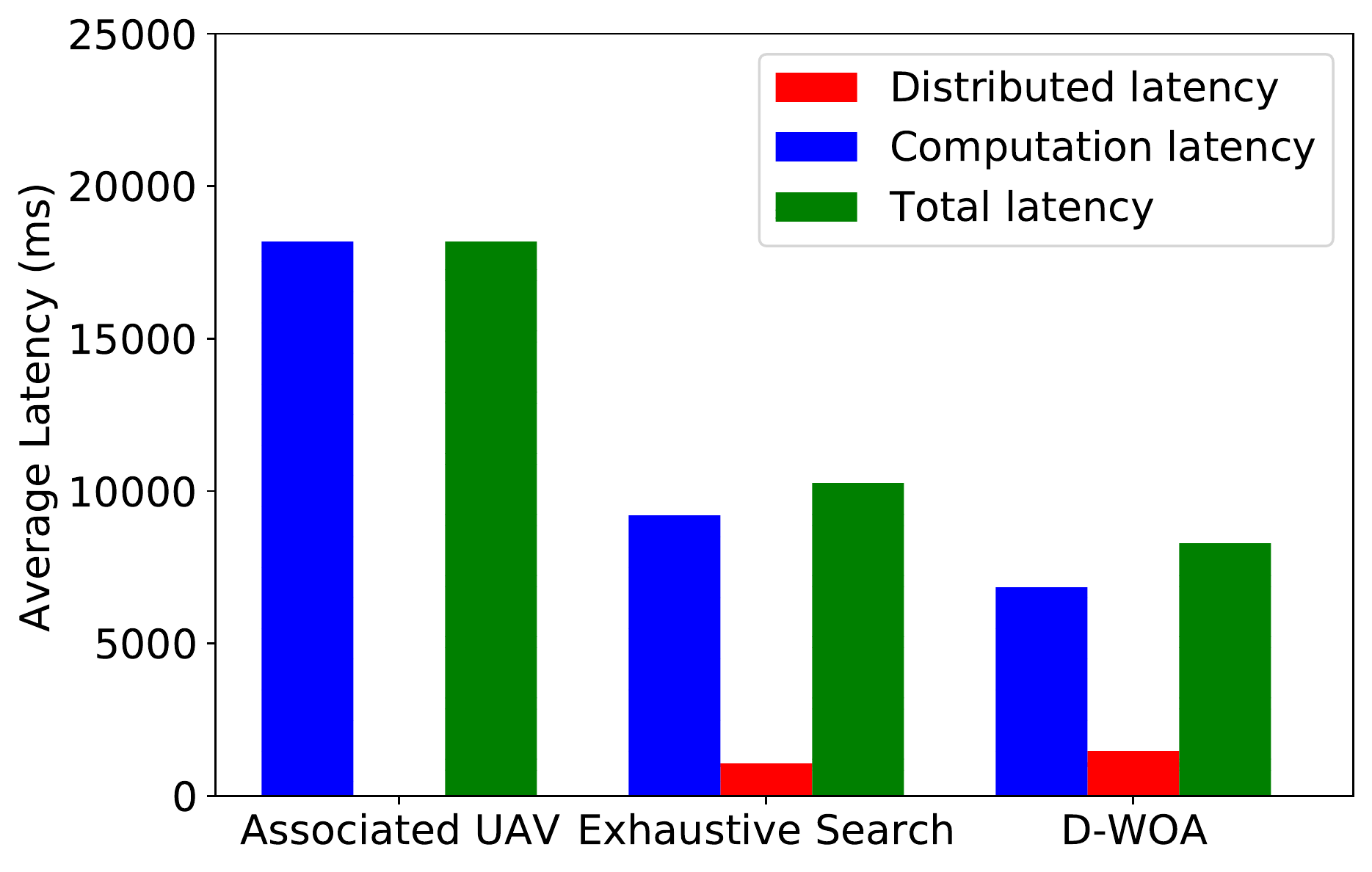}\\
    \caption{Performance comparison under different algorithms.}
    \label{fig:Result2}
\end{figure}

In Fig.~\ref{fig:Result2}, we compare our solution (D-WOA) for the offloading decision problem with associated UAV and exhaustive search approaches under different kinds of latency such as distributed latency,  computation latency, and total latency. As the name it sounds, the computation latency includes the time MEC-enabled UAVs need to execute the sub-tasks. On the other hand, the distributed latency is the time for the associated UAV to distribute the sub-tasks to the other UAVs responsible for executing the sub-tasks. Therefore in the associated UAV solution, the distributed latency is equal to 0. The distributed latency depends on the offloading decision of the algorithm. Finally, the total latency is the sum of two previous latency and also our objective function in (46). In this scenario, total latency under our proposed solution, exhaustive search, and associated UAV scheme are 8,286 ms, 10,248 ms, and 18,186 ms, respectively. Based on the outcomes above, our proposed D-WOA method outperforms other schemes: particularly 54.43\% better than the associated scheme and 19.145\% better than the exhaustive search in terms of total latency. As we can see, the associated UAV scheme, in which there is no cooperation among UAVs, perform poorly compared to the exhaustive search algorithm (ESA) and proposed D-WOA. This outcome indicates the effectiveness of the collaboration of UAVs in the wireless network.

\begin{figure}
    \centering
    \includegraphics[width=.48\textwidth]{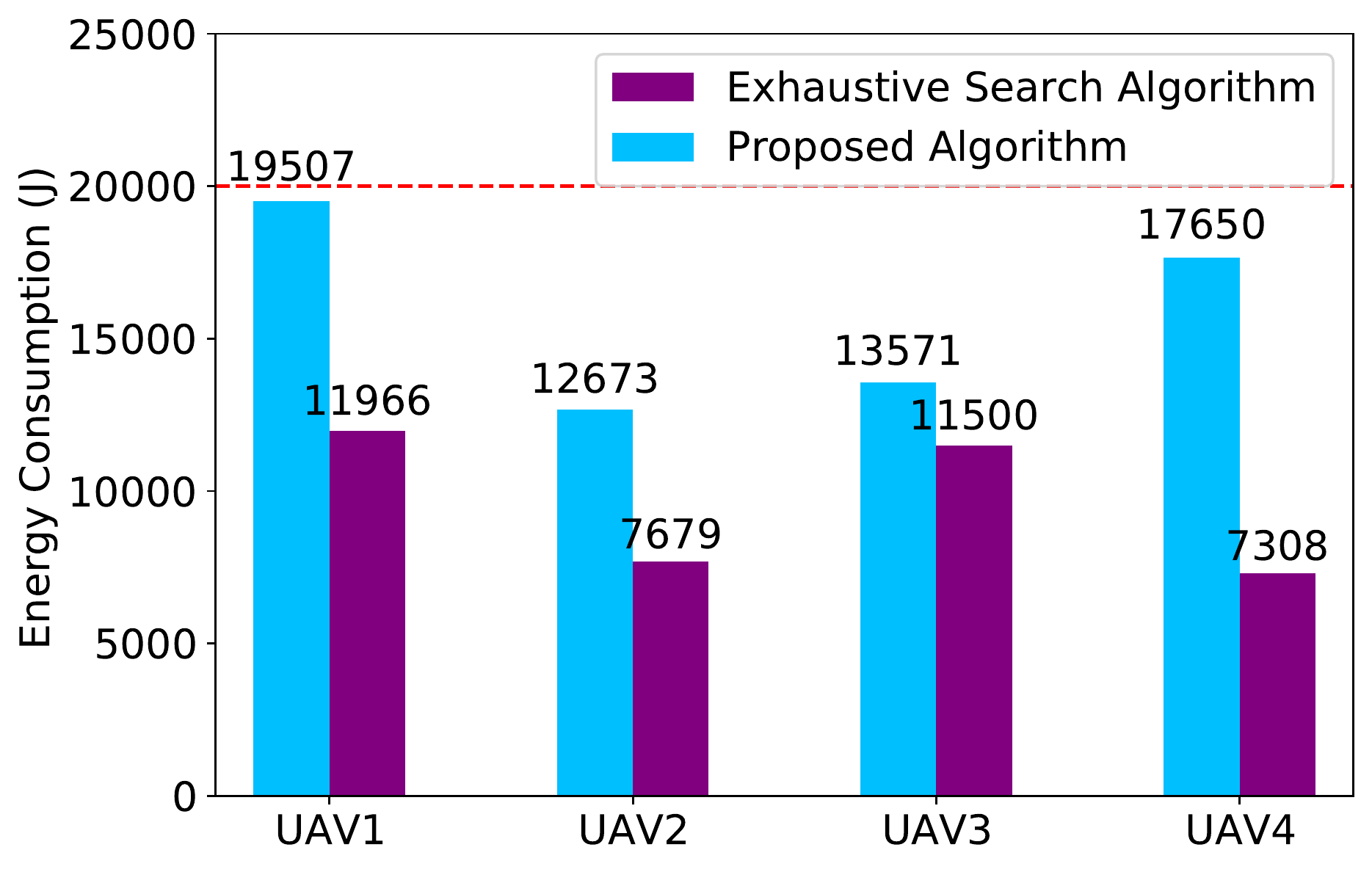}\\
    \caption{Energy consumption by each UAV.}
    \label{fig:Result3}
\end{figure}

Fig.~\ref{fig:Result3} shows the energy consumption of each UAV for ESA and D-WOA. As we can see, the D-WOA forces the UAVs to utilize their available power effectively, while the solution provided by ESA only exploits 50\% of the usable energy. Normally, the maximum energy of a UAV is around $200$ kJ, however, in this simulation, we focus on evaluating the proposed scheme under limited energy so that the maximum energy of a MEC-enabled UAV will be set based on the number of sub-tasks as follows: $E^{\textrm{max}}_{v}=N_{u}$ $\times$ 2,000. As mentioned above, each task of a user has 10 sub-tasks therefore $E^{\textrm{max}}_{v}=$ $20$ kJ . When the number of sub-tasks per task increases, the maximum energy also increases to guarantee a feasible solution. 

\begin{figure}
    \centering
    \includegraphics[width=.48\textwidth]{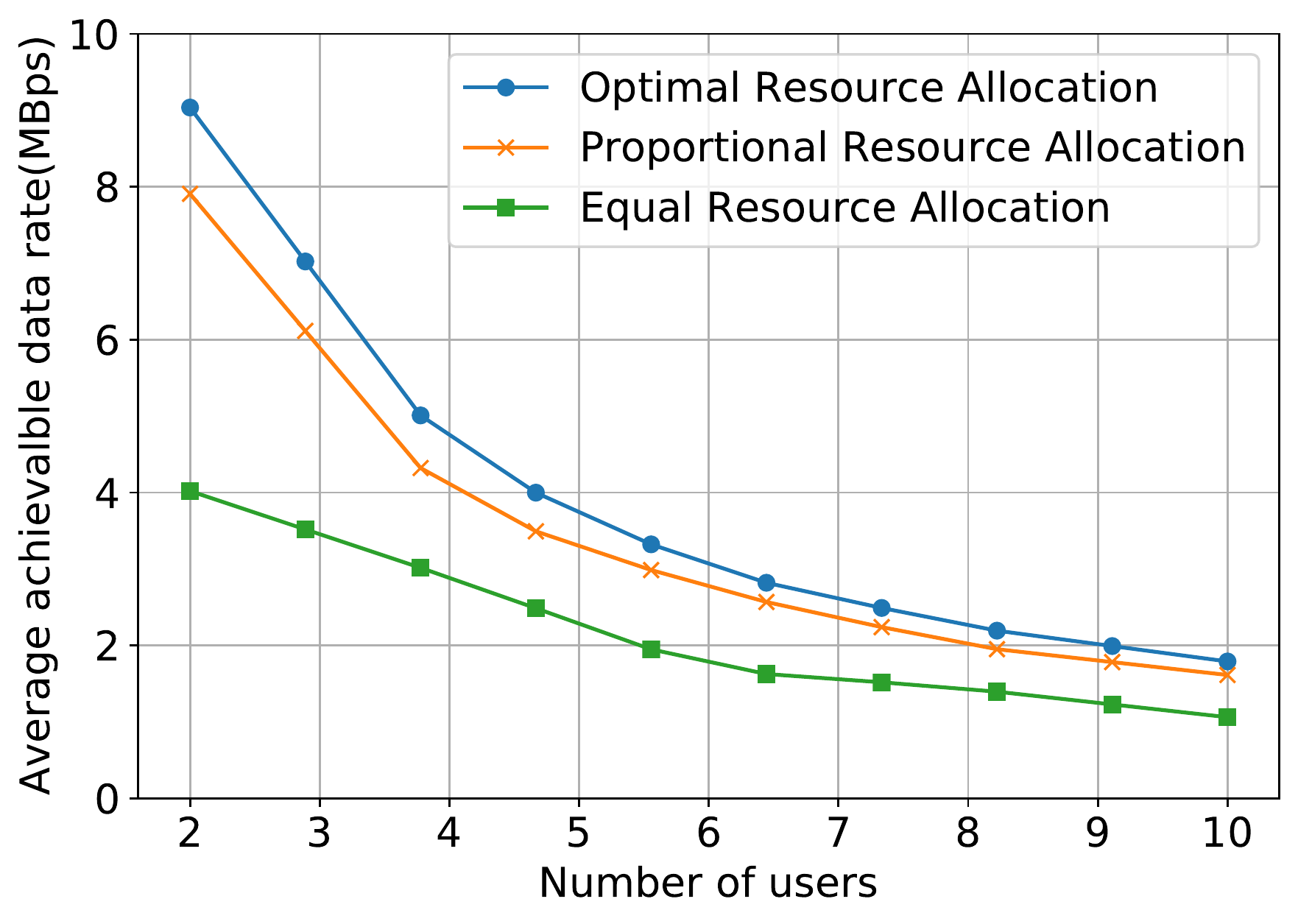}\\
    \caption{Average data rate under different numbers of user.}
    \label{fig:Result4}
\end{figure}

In Fig.~\ref{fig:Result4}, we show the average achievable data rate of the users communicating with its associated UAV in a single subarea, the same way can be applied to other subareas. In the equal resource allocation scheme, all communication resources (i.e., bandwidth) available at the UAV are equally allocated among its associated users within the area, whether they need it or not in a fair manner. While in the proportion resource scheme, the MEC-enabled UAV allocates communication resources to the users based on the network information: the active users and the size of offloading tasks as the following equation:
\begin{equation}
    \beta^{v\rightarrow u}= \frac{H_{u,v}}{ \sum_{q\in\textit{v}} H_{u,v}}B^{v} ,\forall{v} \in \mathcal{V},\forall{u} \in \mathcal{U}_{v}.
\end{equation}
\begin{figure}
    \centering
    \includegraphics[width=.48\textwidth]{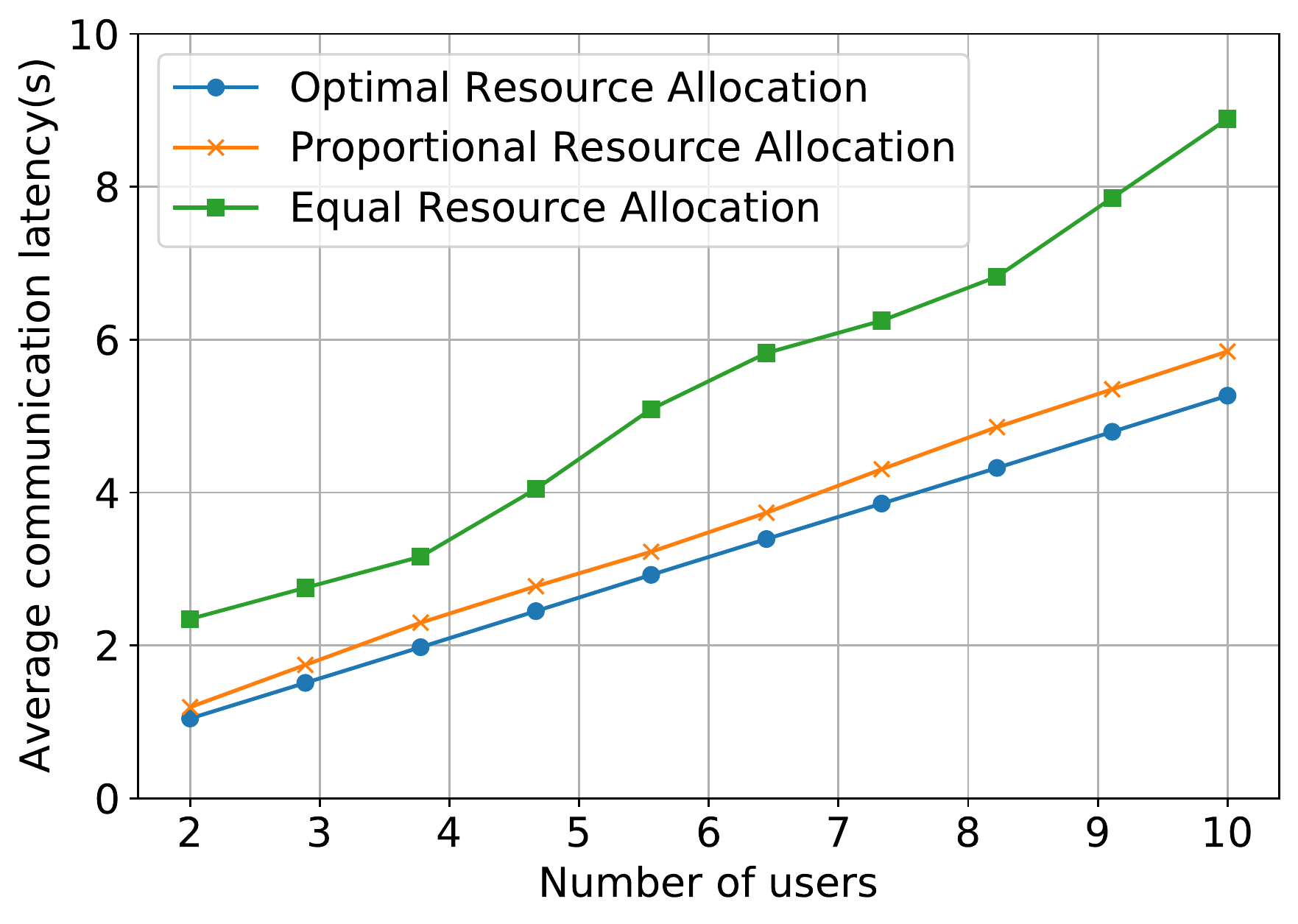}\\
    \caption{The communication latency  number of users.}
    \label{fig:Result5}
\end{figure}

As shown in Fig.~\ref{fig:Result4}, we compare the average uplink data rate of our proposed solution with proportional resource allocation and equal resource allocation schemes under the different numbers of users in the subarea. Generally, when the number of users increases, the average data rate of all schemes decreases. This happens because of the limitation of bandwidth, more users have to share a fixed amount of bandwidth. As we can see in Fig.~\ref{fig:Result4}, our resource allocation solution yields a higher data rate compared to other schemes in every circumstance. Additionally, in Fig.~\ref{fig:Result5}, we display the average communication latency for various network sizes. It is easy to understand why this happens, the communication latency is calculated by the input size of the task divided by the achievable data rate, which means communication latency and achievable data rate are inversely proportional, so when the data rate decreases dramatically (shown in Fig. 5), the communication
latency grows significantly (shown in Fig. 6). Fig. 6 also compares the average latency under different schemes, and
we can see that the optimal resource allocation gives the lowest latency among the three of them, 9.89\% better than the proportional resource allocation approach, 40.74\% better than the equal resource allocation approach when the subarea has 10 mobile users. These percentages are even more when the network has fewer users. With these results, we can conclude that the proposed method for the resource allocation problem outperforms other schemes.

\begin{figure}
    \centering
    \includegraphics[width=.48\textwidth]{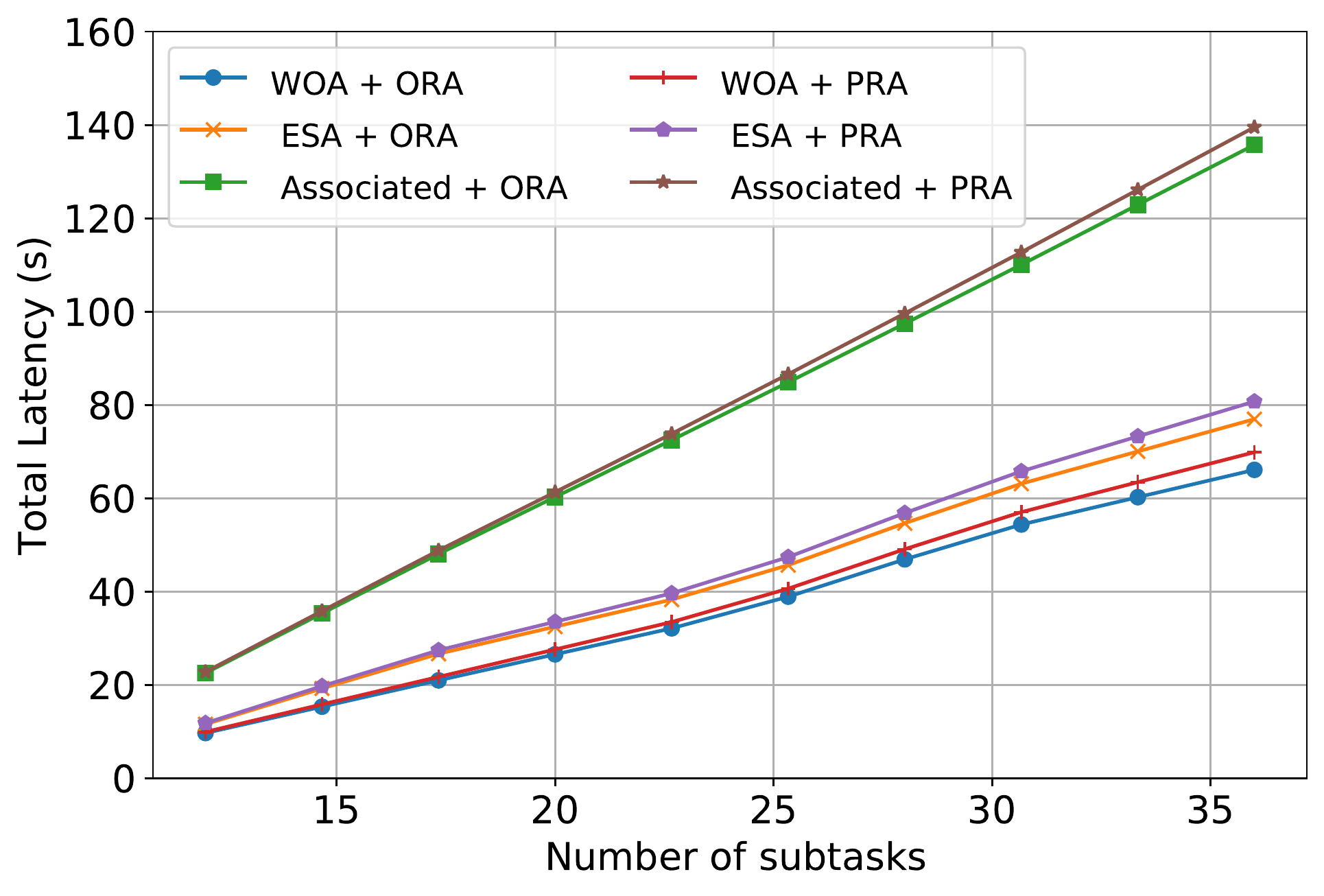}\\
    \caption{Total latency of different algorithm versus the number of sub-tasks (Unlimited Energy).}
    \label{fig:Result6}
\end{figure}

In Fig.~\ref{fig:Result6}, we compare the result of our proposal for the joint offloading decision and resource allocation problem with other benchmarks in terms of total latency. Our approach, as far as we can tell, outperforms any combination of exhaustive search (ESA), associated UAVs, and proportional resource allocation (PRA). When the number of sub-tasks in each task increases, the total latency of all algorithms also increases rapidly. However, their slopes differ: associated schemes have the steepest slope, followed by the ESA scheme and our proposed solution. This indicates that our method works effectively in a large dimension space (more sub-tasks, more offloading variables), while the other schemes perform poorly. For example, when there are 36 sub-tasks for a single task, the total latency is 66.080 s (WOA+ORA), 76.952 s (ESA+ORA), and 135.739 s (Associate + ORA). These results indicate our proposed approach is 14.13\% better than the greedy approach and 51.32\% compared with the associate scheme. The explanation for the improvement is that the associated UAVs scheme executes the tasks alone. There are cases when two sub-tasks can be executed in parallel, but the associated UAVs only execute one of them or divide the computing resources to execute both of them at the same time. In either case, computing latency increases. For the ESA case, the collaboration among UAVs lowers the latency, but it is extremely hard to find the best answer among all feasible solutions without an appropriate searching strategy. The integration of different searching mechanisms in D-WOA has proved its effectiveness by providing the lowest latency in the simulation.
\begin{figure}
    \centering
    \includegraphics[width=.48\textwidth]{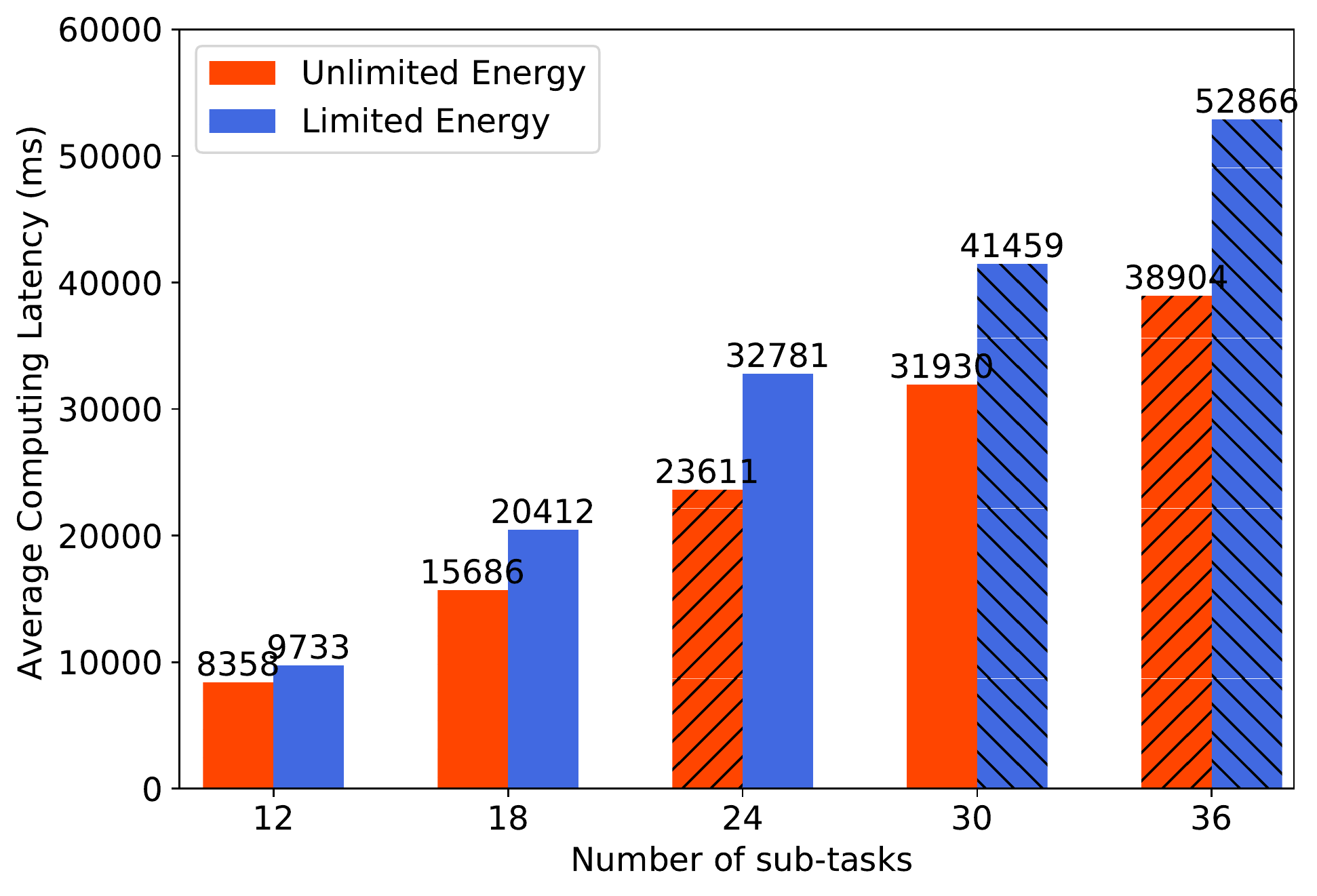}\\
    \caption{The average computing latency between limited \& unlimited case under different numbers of sub-tasks.}
    \label{fig:Result7}
\end{figure}

In Fig.~\ref{fig:Result7}, we compare the average computing latency of D-WOA in two cases: unlimited and limited energy. From the latency result, we can easily see the effect of energy constraint here: the infinite energy case achieves better latency than the limited case. The gap between the two schemes gets wider with the increase of sub-tasks. When the energy of UAVs is limited, the UAVs have to lower their computing resources to execute the task, as (25) leads to reducing the computing energy. With the lower computing resource and the inversely proportional relationship between computing latency and computing resources in (22), the execution time of the task increases, which is obviously appropriate.

\begin{figure}
    \centering
    \includegraphics[width=.48\textwidth]{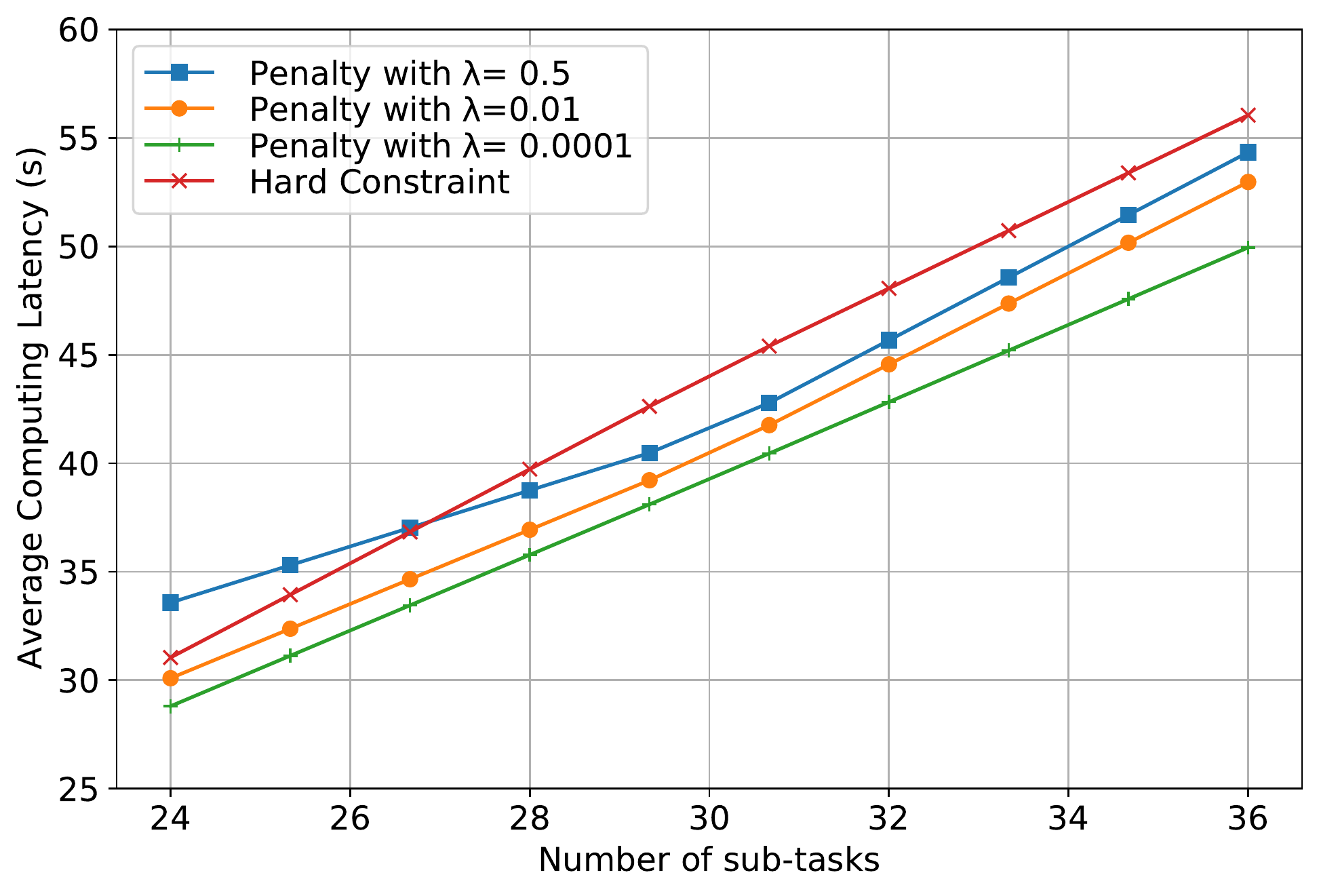}\\
    \caption{The average computing latency among different penalty factor of sub-tasks.}
    \label{fig:Result8}
\end{figure}

In Fig.~\ref{fig:Result8}, we do experiments to show the penalty factor's impact on computing latency. Various values of impact factors have been evaluated, however, we only show some worth notice value in the figure. Before going into detail, we need to explain the Hard Constraint case: if the offloading decision violates the energy constraint, we will omit that solution by assigning a large number to the objective function. As we observe from Fig.~\ref{fig:Result8}, the penalty technique offers a better solution than the hard constraint scheme. The penalty factor $\lambda$ is small, which implies that the searching agents are encouraged to explore more promising regions to find the best solution, even if it violates the constraint. We get a lower computing latency when we decrease the penalty factor from 0.5 to 0.01. However, the smaller the better is not always right; in this example also, the scheme with the penalty factor $\lambda$=0.0001 returns an infeasible solution-the objective function is not penalized enough. To ensure the algorithm returns a feasible solution, the penalty factor has to stay in the range [0.01 – 0.5].

\section{Conclusion}\label{Conclusion}
In this article, we have studied the topology of tasks in the real world: to be more specific, how the dependency has effects on the start time of successor sub-tasks and the finish time of the task. We then considered the scheme where dependency tasks from mobile users are offloaded to collaborative MEC-enabled UAVs networks. Then we formulated an optimization problem with the goal of minimizing the average latency experienced by mobile users by optimizing the offloading decision for each sub-task of users' tasks and communication resource allocation. The designed problem was mixed-integer, non-linear, and non-convex when considering it as a whole. Therefore, to solve the problem in polynomial time and make it easier, we decomposed the formulated problem into two sub-problems: the offloading decision problem and the communication resource allocation problem. Then, we proposed an appropriate solution: a heuristic solution, D-WOA, to acquire the optimal offloading matrix for offloading decision problem. Next, we used SCS solver in the library CVXPY in order to solve the communication resource allocation problem. Finally, we carried out extensive simulations to demonstrate the superior performance of our algorithm compared to existing benchmark schemes in terms of the total delay encountered by users. Our proposed scheme achieved the lowest latency and outperformed other benchmark schemes in literature.

% or


\appendices

\ifCLASSOPTIONcaptionsoff
  \newpage
\fi

\bibliographystyle{IEEEtran}
\bibliography{mybib}
\end{document}